\documentclass[aps,pre,reprint,superscriptaddress,amsmath,amssymb,byrevtex,floatfix]{revtex4-2}

\usepackage{graphicx}
\usepackage{siunitx}
\usepackage[colorlinks=true, allcolors=blue]{hyperref}
\usepackage{amsmath}
\usepackage[normalem]{ulem}
\usepackage{mathtools}
\usepackage{bm}

\begin{document}

\title{Dissolution-driven propulsion of floating solids}

\author{Martin Chaigne} \email{martin.chaigne@u-paris.fr} \affiliation{Laboratoire Mati\`ere et Syst\`emes Complexes, Université Paris Cité, CNRS (UMR 7057), F-75013 Paris, France}
\author{Michael Berhanu} \email{michael.berhanu@u-paris.fr} \affiliation{Laboratoire Mati\`ere et Syst\`emes Complexes, Université Paris Cité, CNRS (UMR 7057), F-75013 Paris, France}
\author{Arshad Kudrolli} \email{akudrolli@clarku.edu} \affiliation{Department of Physics, Clark University, Worcester, Massachusetts 01610, USA}

\date{\today}

\begin{abstract}
We show that unconstrained asymmetric dissolving solids floating in a fluid can move rectilinearly as a result of attached density currents which occur along their inclined surfaces. Solids in the form of boats composed of centimeter-scale sugar and salt slabs attached to a buoy are observed to move rapidly in water with speeds up to 5 mm/s determined by the inclination angle and orientation of the dissolving surfaces. While symmetric boats drift slowly, asymmetric boats are observed to accelerate rapidly along a line before reaching a terminal velocity when their drag matches the thrust generated by dissolution. By visualizing the flow around the body, we show that the boat velocity is always directed opposite to the horizontal component of the density current. We derive the thrust acting on the body from its measured kinematics, and show that the propulsion mechanism is consistent with the unbalanced momentum generated by the attached density current. We obtain an analytical formula for the body speed depending on geometry and material properties, and show that it captures the observed trends reasonably. Our analysis shows that the gravity current sets the scale of the body speed consistent with our observations, and we estimate that speeds can grow slowly as the cube-root of the length of the inclined dissolving surface.  The dynamics of dissolving solids demonstrated here applies equally well to solids undergoing phase change, and may enhance the drift of melting icebergs, besides unraveling a primal strategy by which to achieve locomotion in active matter. 
\end{abstract}

\maketitle

%\section{Introduction}
Self-propulsion by converting stored energy into mechanical motion is at the heart of active matter~\cite{Marchetti2013}. The motion may occur through a chemical reaction enabling ciliary beating in micro-swimmers~\cite{Brennen1977,Lauga2009,Elgeti2015}, or by generating ballistic motion at molecular scales by catalytic boosts of enzymes~\cite{Jee2018}. At the granular scale, directed motion can be observed on vibrating substrates via spontaneous symmetry breaking~\cite{dorbolo2005} or more robustly with polar grains~\cite{kudrolli2008}. Self-propulsion can also be  created without mechanical action.  For example, asymmetric particles which catalyze a chemical reaction in the fluid can break mechanical equilibrium~\cite{Golestanian2005}. Chemical or temperature gradients can also induce variations of the surface tension on droplets generating propulsion due to the Marangoni effect~\cite{Izri2014,Maas2016,Ryazantsev2017,Reichert2021}, and Leidenfrost droplets can experience propulsion caused by interactions with the substrate~\cite{Linke2006,Lagubeau2011,Gauthier2019}. 

Density currents resulting from spatial variations in fluid density constitute another possible mechanism for self-propulsion. This mechanism, which is not restricted to a free-surface as in the Marangoni effect and is not limited to microscopic scales, has received limited attention. Passive asymmetric solids  floating in density stratified fluids have been reported to experience thrust leading them to move slowly with speeds of a few microns per second~\cite{Allshouse2010}. Faster transport can occur if the body itself generates density variation in the surrounding fluid as demonstrated by Mercier, et al.~\cite{Mercier2014} with a floating asymmetric solid with an embedded local heat source that generates thermal convection. However, convective flow  can also occur without an added heat-source through the progressive phase change of a solid immersed in a fluid. Gravity currents due to solute concentration gradients have been studied to understand shapes changes in dissolving solids~\cite{Sullivan96,Wykes2018,Philippi2019,Cohen2020,Huang2020}, but whether these flows lead to self-propulsion of the solid itself was unprobed. While auto-rotation of floating ice disks caused by melting has been noted~\cite{Dorbolo2016}, net translation was not noted in these studies. More recently, dissolving colloidal particles driven by Brownian motion have been considered theoretically and predicted to undergo stochastic dynamics~\cite{Chamolly2019}, but any effect of convective flow of the surrounding fluid as a result of the dissolution was not examined. 

Here, we show that an unconstrained asymmetric dissolving solid can propel itself rectilinearly because of the thrust generated as a reaction to the unbalanced momentum of the solute-rich density current which develops along its inclined surfaces.  While we focus here on the case of dissolution because it does not have the complexity associated with temperature gradients and phase change, the elucidated propulsion mechanism applies equally well to asymmetric bodies undergoing melting. We further discuss the implication for active transport of floating ice as a result of melting or freezing, relative to their advection due to wind and ocean currents~\cite{Mountain1980,Anderson2016,Feltham2008,Alberello2020}. 

\newpage
\section{Results}
\begin{figure*}[t]
   \centering
   \includegraphics[width=1\textwidth]{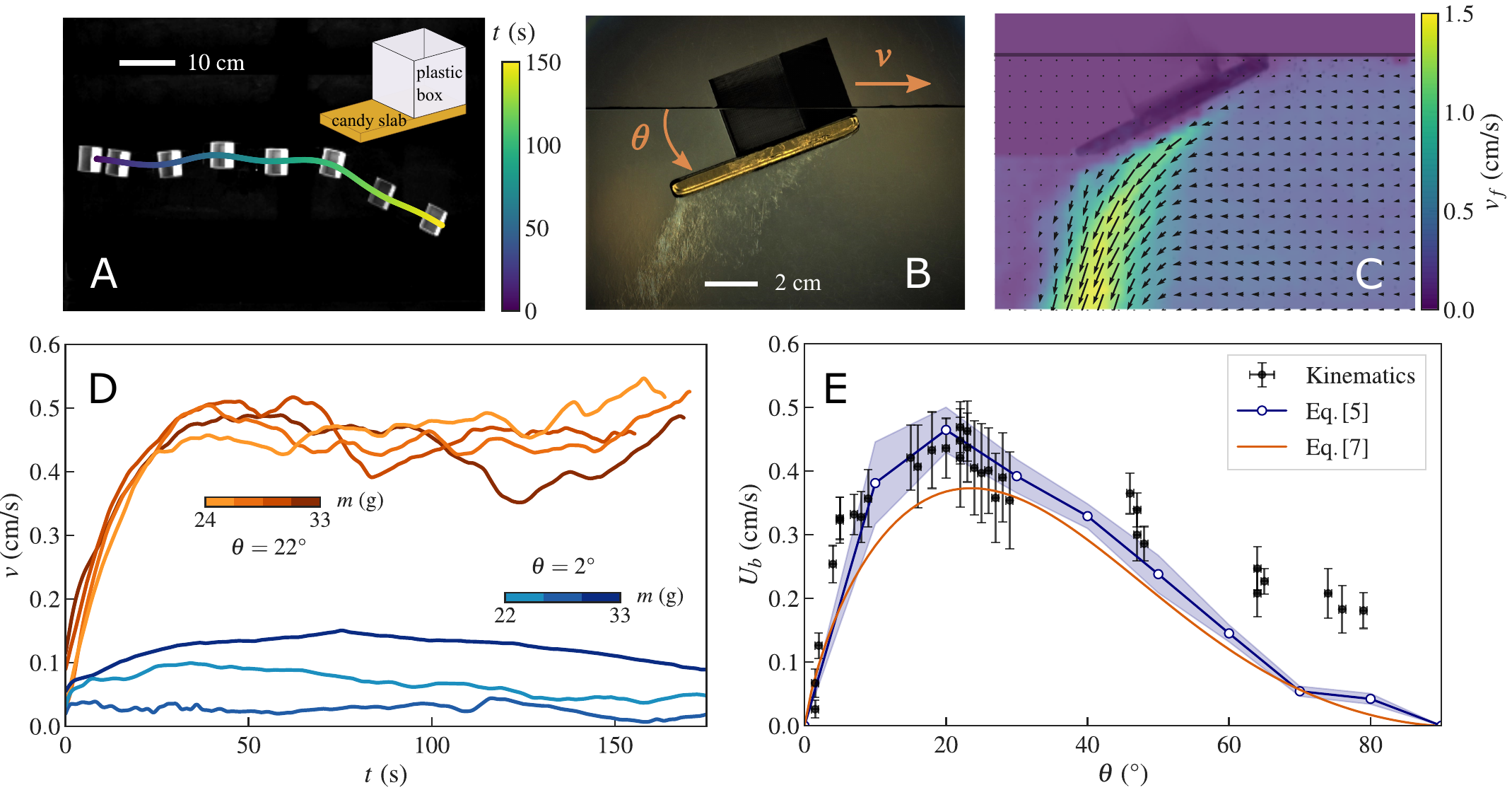}
    \caption{\label{fig:image} Kinematics of the candy boats. A: Superimposed images at $\SI{20}{\second}$ time intervals show the motion of a candy boat in a large tank of water (top view). Inset: Schematic of a candy boat. B: An image of a candy boat moving in water obtained with shadowgraph technique (side view). Solute-rich plumes  are visible descending below the dissolving boat. C: The fluid velocity $v_f$ in the body frame of reference measured with PIV in the mid-vertical cross-section of the boat superimposed on an image of the system  ($\theta = \SI{26}{\degree} \pm 1^\circ$). The magnitude and direction of $v_f$ averaged over 15 seconds are shown with arrows and color map.  D: The evolution of boat velocity $v$ corresponding to consecutive launches with two different boats: $\theta = 22^\circ \pm 1^\circ$ (orange lines) and $\theta = 2^\circ \pm 1^\circ$ (blue lines). The mass $m$ decreases with each trial and is shown by color bar for each boat. E: The measured boat speed $U_b$ ($\bullet$) varies significantly with $\theta$, and is well described by estimates obtained by performing a momentum balance analysis (Eq.~[\ref{eq:Ubthe}]) with measured flow fields in Section~\ref{sec:propmech}{, and by the functional analysis which leads to Eq.~[\ref{eq:Uvp}] combined with Eq.~[\ref{eq:vp}] in Section~\ref{sec:funcUb}}. The blue shade corresponds to the estimated error in determining $F_p^{MB}$. Systematic deviations are observed at higher $\theta$ because only the contribution of the flow below the dissolving slab is considered for simplicity of analysis. }
\end{figure*}

\subsection{Design and observations}
We construct dissolving bodies in the form of boats which float asymmetrically in water by attaching a solid $7.5 \times 4.0 \times 0.5$\,cm$^3$ rectangular candy slab to the bottom of a hollow plastic box which acts as a buoy.  A schematic of such a candy boat is shown in the Inset to Fig.~\ref{fig:image}A and further details on its fabrication process can be found in the Methods section. When the composite boat floats in water, the placement of the buoy relative to the center of the heavier candy slab determines the inclination angle $\theta$ of the dissolving surfaces as shown in Fig.~\ref{fig:image}B. In all, a set of 9 boats with mass $m$ of approximately $33$\,g each were constructed with identical dissolving slabs and buoys to vary $\theta$ between $0^\circ$ and $80^\circ$, and to study the effect of breaking fore-aft symmetry on their dynamics.  

Figure~\ref{fig:image}A and SI Movie~S1 show a time-sequence of a dissolving solid boat  corresponding to $\theta = 22^\circ$ moving in fresh water after it is released from rest in a large tank.   We observe from the superimposed trajectory of the candy boat that it moves essentially rectilinearly while achieving speeds of order 5 mm/s. By contrast, we observe that the boat drifts slowly with speeds less than 0.7\,mm/s if the plastic hull is placed centrally which results in $\theta \approx 2^\circ$ (see SI Movie~S2). Thus, symmetry breaking due to the inclination of the dissolving body is important to the observed rapid rectilinear motion. 

We further visualize the solvent bath to examine the solute-rich fluid which occurs around the dissolving body. The images shown in Fig.~\ref{fig:image}B and SI Movie~S3 are taken with shadowgraph technique which magnifies refractive index variations caused by the solute concentration. Turbulent plumes are observed to emerge directed behind and below the dissolving surface as the boat moves forward, but the area near the air-water interface around the boat does not show any refraction due to the presence of solute which may signal surface tension gradients. We further visualize the fluid motion with Particle Image Velocimetry (PIV) by adding fluorescent tracers as discussed in SI Section~1. PIV measurements. Fig.~\ref{fig:image}C shows the corresponding flow field $v_f$ obtained with PIV in a vertical cross sectional plane. To our knowledge, the convection flow below a dissolving body has not been quantitatively characterized previously. The fluid can be observed to accelerate rapidly below the inclined dissolving surface before detaching and flowing downwards with the greatest velocities directed below and behind the dissolving surface. Because the solute-rich fluid descends rapidly to the bottom of the tank, the fluid near the surface remains solute-free, since the fluid is drawn in apparently faster than the time scale over which the solute can diffuse out around the dissolving solid.  Thus, the boat is seen to move opposite to the direction of the density current which flows towards the back, while moving down due to the action of gravity.  

By adding weights to the plastic box, we submerged the entire body to a depth of 4\,cm, turning it into a submarine. To maintain a constant immersion depth, a stratified bath was prepared with a fresh water layer lying over a denser salt water layer (for more details see SI Section 2. Fully submerged body). We observe that the boat moves similarly as shown in SI Movie~S4 with speeds up to 1\,mm/s, roughly the same order of magnitude as when floating at the air-water interface. This observation further confirms that the propulsion mechanism at play is different from Marangoni flows which can propel dissolving solids such as camphor boats due to surface tension gradients~\cite{Nakata1997,Nagayama2004,Biswas2020}.  

We also performed measurements with dissolving boats where the sucrose block was replaced with a salt (NaCl) block with dimensions 4.1\,cm $\times$ 2.3\,cm $\times$ 0.6\,cm to test whether the propulsion can be observed with other dissolving materials.  The same qualitative behavior is observed with {$U_b \approx 4.2$\,mm/s} when $\theta = 40^\circ$ (see SI Section~3.~Effect of Dissolving Material). Unlike sugar-water solutions where viscosity can vary several orders of magnitude when the saturation limit is reached, the viscosity of saturated salt solution is only higher by a factor 2  compared with water~\cite{Handbook}. Thus, the large viscosity variations specific to sugar solutions do not play an appreciable role in the propulsion.

{To further check the robustness of the propulsion mechanism, and its persistence in multi-body environments, we performed experiments with two candy boats with the same length $L = 7.5$\,cm, and $\theta = 33^o \pm 2^o$. As shown in SI Movie~S5, when the boats are moving in a row in the same direction, the following boat catches up with the leading boat, and the boats self-assemble to move forward in tandem. While approaching from opposite directions, the boats come in contact, pair up, and then spin around each other as shown in SI Movie~S6 and SI Movie~S7, with speeds that depend on their relative contact position (see SI Section 4. Boat Pairs). Thus, we find that while capillary interactions, important over the scale of a centimeter or less~\cite{Vella2005,Dalbe2011},  cause these bodies floating at the surface to stay in contact, it is clear that such boats can move collectively and can show further rich phenomena. }

\subsection{Kinematics}
Figure~\ref{fig:image}D shows the measured boat velocity $v$ obtained from the displacement of its center of mass position over a time interval of $\SI{0.2}{\second}$ as a function of time $t$ corresponding to $\theta = 22^\circ$. Four consecutive launches of the same boat are shown to illustrate the robust features of the kinematics. We observe that $v$ increases smoothly before the turbulent nature of the plumes and rudderless nature of the boat causes its velocity to fluctuate somewhat randomly. The total mass of the boat is observed to decrease by approximately 10\% over the course of the entire trial. Since $\theta$ depends on the relative mass and location of the buoy and the heavier dissolving block, it can evolve slowly over time in principle with dissolution. However, the measured $\theta$ was observed to be constant to within $1^\circ$ in these examples, and no systematic variation from one launch to the next can be observed. We obtain the boat cruising speed $U_b \approx 4.5$\,mm/s by averaging over the time after the boat stops accelerating forward (about $t = 40$\,s here), and while it maintains a more or less constant speed. These observed speeds are faster by an order of magnitude compared with previous demonstrations of temperature-gradient driven convective transport in immersed bodies~\cite{Mercier2014}, and more than two orders of magnitude faster than diffusion-driven transport in stratified fluids~\cite{Allshouse2010}.

We investigate $U_b$ further as a function of the inclination of the dissolving surfaces using the nine different candy boats constructed with different buoy locations. Figure~\ref{fig:image}E shows the mean $U_b$  and $\theta$ along with their standard deviations for a given trail. We observe that $U_b$ increases rapidly reaching a peak at $\theta \approx 22^\circ$, before decreasing as $\theta$ increases toward $80^\circ$. The overall increase and then decrease of speed with $\theta$ is consistent with the rectangular geometry of a dissolving block which would be symmetric about the vertical axis, when $\theta = 0^\circ$ and $90^\circ$. Thus, the significant dependence of speed on $\theta$ highlights the importance of orientation of the dissolving surfaces on the propulsion of the dissolving bodies in our experiments.

\subsection{Thrust and Drag}  
\begin{figure}[t]
\centering
\includegraphics[width=1\columnwidth]{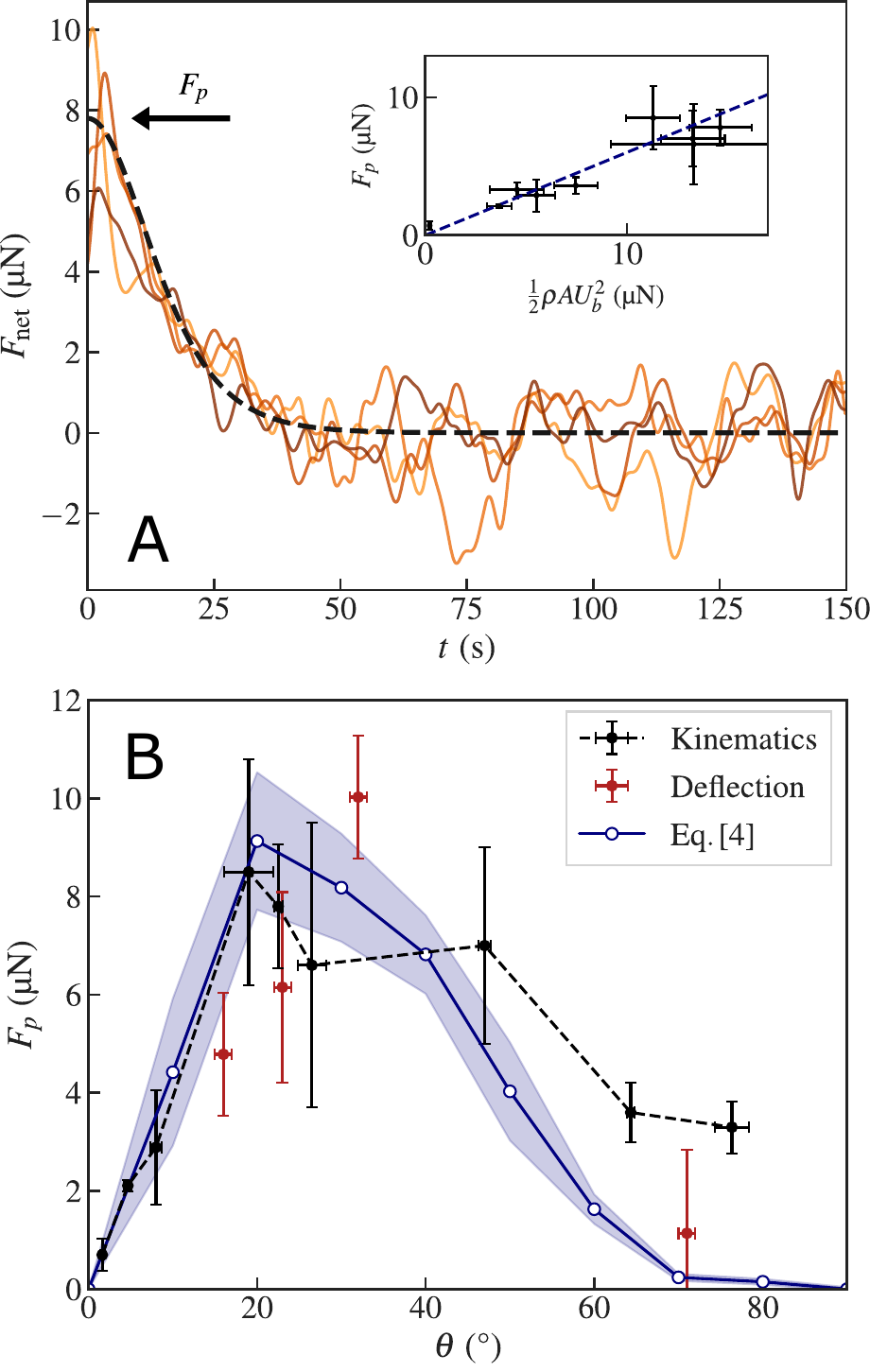}
\caption{\label{fig:Force} Forces acting on a moving dissolving body. A: The net force $F_{\mathrm{net}}$ decays to zero as the drag increases with speed and balances the thrust $F_p$. The dashed line corresponds to Eq.~[S2] assuming $v^2$-drag (see SI text 4). Inset: $F_{p}$ versus $F_{d}/C_d = \frac{1}{2} \rho L_A W U_b^2$. The data is described by a linear fit corresponding to quadratic-drag with $C_d = 0.6$. B: The thrust $F_p$ obtained using kinematic measurements, deflection measurements, and calculated using the momentum balance condition given by Eq.~[\ref{eq:Fp}], and its estimated error (blue shade).}
\end{figure}

We examine the forces acting on the floating body as it dissolves, toward understanding its kinematics. Because the identification of distinct portions of the body over which the propulsive and the retarding forces apply is complex, we assume that the force exerted by the fluid on the boat is written as a constant term $F_p$, which we call “thrust”, minus a term increasing with the velocity $F_d$, which we call “drag”. Such a decomposition is widely used for example to model the kinematics of swimming bodies in the inertial regime~\cite{Gazzola2014,Gazzola2015,VanBuren2018}, when the thrust and the drag result from a pressure field. The boat accelerates due to these unbalanced forces acting on it toward the bow along its symmetry axis. Neglecting the change in mass of the body, the net force $F_{\mathrm{net}} = F_p - F_d$ is proportional to the acceleration of the body according to: $F_{\mathrm{net}} = m\,{\rm d}v/{\rm d}t$, where we ignore the added mass effect which may arise because moving a body requires displacing the fluid in which it is immersed~\cite{GuyonHulinPetit,BrennenBook,BrennenReview}. We plot $F_{\mathrm{net}}=m\,{\rm d}v/{\rm d}t$ as a function of time in Fig.~\ref{fig:Force}A, and observe that it is highest close to $t=0$\,s, when $v \approx 0$\,mm/s. $F_{\mathrm{net}}$ then decreases rapidly toward zero as drag increases and balances the thrust. When $v \rightarrow 0$, we have $F_d \rightarrow 0$ and $F_{\mathrm{net}} \rightarrow F_p$. We plot $F_p$ as a function of $\theta$ in Fig.~\ref{fig:Force}B, and observe that it increases and decreases with $\theta$ following the trends in $U_b$ with $\theta$. We performed further complementary measurements of $F_p$ by obstructing the boat's forward motion using a long thin rod and measuring its deflection (see Fig.~\ref{fig:Force}B). The measured thrusts using the two methods are in overall agreement, confirming that the added mass effect can be neglected at least at low to moderate $\theta$. 

If the thrust does not change with $v$, we have $F_d = F_p$ when the boat moves with speed $U_b$. The drag can be written in general as  $F_{d} = \frac{1}{2}C_d \rho_f A U_b^2$, where $\rho_f = 997$\,g\,L$^{-1}$ is the density of water, $A$ is the projected area along the direction of the boat's velocity, and $C_d$ is the drag coefficient. We have $A = L_A\, W$, where $W$ is the width and $L_A$ is the projected length measured for each boat from a side view image corresponding to the projected length of its immersed part on the vertical axis  (see SI Section~5.~Projected Boat Length). The Reynolds Number over the scale of the boat can be estimated as $Re = U_b\, L/\nu \approx 340$, since $L = 7.5$ cm, $U_b\approx 4.5$ mm/s and $\nu =1.00\times 10^{-6}$ m$^2$ s$^{-1}$, the kinematic viscosity of water at 20$^\circ$C. Since $Re \gg 1$, the drag can be expected to scale quadratically with speed. We plot $F_p$ versus $F_d/C_d = \frac{1}{2} \rho L_A W U_b^2$ in the inset of Fig.~\ref{fig:Force}A, where each point represents averaged values over trials of the same boat. The data is observed to be well described by a linear fit, with a slope corresponding to $C_d = 0.6$, which is reasonable considering $C_d$ for a non-streamlined body~\cite{GuyonHulinPetit}.

We calculate an analytical expression for $F_{\mathrm{net}}$ over time assuming thrust $F_p$ independent of $v$, and $F_d = \frac{1}{2}C_d \rho A v^2$ in SI Section~6.~Net Force Evolution. Plotting $F_{\mathrm{net}}$ using the measured ratio $C_d = 0.6$ in Fig.~\ref{fig:Force}A (dotted line), we find good agreement with the time scales over which the boat accelerates. This agreement validates our assumption that the thrust generated by the dissolution is essentially independent of the speed of the boat, and depends only on the angle of inclination of the dissolving surface. Even though it is not obvious in general, the agreement also confirms \textit{a posteriori} the decomposition of forces into thrust and drag on the same surface. Thus, we find that a dissolving body released from rest accelerates and reaches a cruising velocity as its drag increases and matches the thrust corresponding to its geometry. 

\begin{table}
\centering
\begin{tabular}{cccc}
\hline
Dissolving Surface & Speed $U_b$ (mm/s) & Area (cm$^2$) & $\dot{h}$ (mm/s)\\
\hline
Bottom and Top & $3.3 \pm 0.2$ & 48 & - \\
Bottom &  $2.6 \pm 0.2$ & 30 & $2.41\times10^{-3}$\\
Top & $0.5 \pm 0.2$ & 18 & $0.87\times10^{-3}$ \\
\hline
\end{tabular}
\caption{\label{tab:speeds} Measured mean speed $U_b$ and its standard deviation corresponding to the exposed dissolving surfaces of the dissolving slab ($\theta = 45^o$). The difference in surface area and dissolution rate can account for the relative contribution of the bottom and top surfaces on $U_b$. }
\end{table}

\subsection{Effect of dissolving surface orientation}
We gauge the relative effect of the upward and downward facing surfaces of the dissolving slab on the propulsion speed by adding a thin plastic wing to the side of the plastic box such that the dissolving slab is covered from above or below. The measured speeds are given in Table~\ref{tab:speeds} corresponding to $\theta = 45^\circ$. While the greatest speed is achieved when both the top and bottom surfaces of the slab are allowed to dissolve, $U_b$ is about 5 times greater when the bottom surface is exposed compared to the top surface. The presence of the buoy screens part of the slab resulting in the area of the bottom surface being nearly 2 times larger than the top surface. Thus, while part of the greater contribution of the bottom surface is due to its larger surface area, bottom facing surfaces dissolve faster than top facing surfaces~\cite{Sharma2021}.

\begin{figure}
\centering
\includegraphics[width=1\columnwidth]{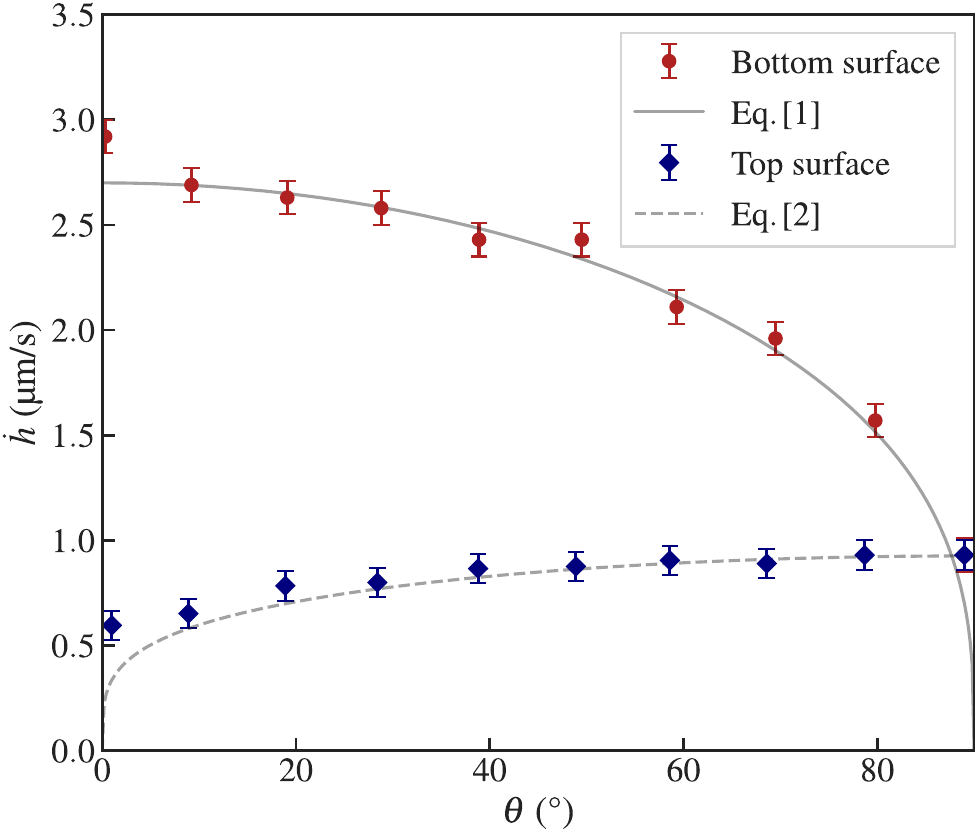}
\caption{\label{fig:erosionrate} The measured recession rate $\dot{h}$ along the top and bottom surface of the dissolving block as a function of its inclination angle $\theta$. Grey lines: Calculated recession rates at the bottom surface (Eq.~[\ref{eq:hdot}], solid line), where a density inversion instability leads to faster $\dot{h}$, with $Ra_c=1101$ and $\nu_i = \SI{2.5e-4}{\square\meter\per\second}$, and at the top surface (Eq.~[\ref{eq:hdottop}], dashed line) with fitting parameter $B=\SI{1.15e-4}{\centi\meter\tothe{5/4}\per\second}$~\cite{Pegler2020}. Greater boat speeds are observed when the bottom surface dissolves compared to top surface at the same inclination angle.}
\end{figure}

To ascertain the effect of inclination on the dissolution of the body and its effect on $U_b$, we measure the location and evolution of the dissolving surface at various $\theta$. Figure~\ref{fig:erosionrate} shows the recession rate $\dot{h}$ of the dissolving surface as a function of its orientation. Because of the solutal Rayleigh-B\'enard instability, $\dot{h}$ is greatest when $\theta = 0^\circ$, but decreases as $\theta$ increases when the dissolving surface faces down. 

The recession rate $\dot{h}$ can be calculated based on material properties and the inclination of the dissolving surface. In the presence of solutal convection~\cite{Sharma2021}, $\dot{h}=\dfrac{D\,\rho_{sat}\,c_{sat}}{\rho_s\,\delta_c\,(\rho_{sat}-c_{sat})}$, where $\rho_s$ is the density of the dissolving solid, $\rho_{sat}$ is the saturation density of the solute, $c_{sat}$ is the saturation concentration, $D$ is the diffusion coefficient, and $\delta_c$ is the concentrated solute boundary layer thickness (see SI Section~7.~Surface Recession Rates). When the surface is oriented downward, the boundary layer is subject to a Rayleigh-Bénard instability. In the quasi-stationary regime, $\delta_c$ is set by the 
critical Rayleigh number $Ra_c$~\cite{Sullivan96,Wykes2018,Philippi2019}.  The effect of the inclination can be taken into account by considering the projection of the gravity perpendicular to this layer~\cite{Cohen2020}. Then,  
%\begin{equation}
$\delta_c=\left(\dfrac{Ra_c\,\nu_i\,\rho_f \,D}{g \cos \theta \, (\rho_{sat}-\rho_f)}  \right)^{1/3}$,
%\label{eq:deltac} 
%\end{equation}
where $Ra_c$ can be approximated as 1101 in case of a Rayleigh-B\'enard instability in a layer with mixed slip and non-slip boundary conditions~\cite{Chandrasekhar}, $\rho_f$ is the density of the far field liquid and $\nu_i$ is the average viscosity of the boundary layer.  
Thus, the recession rate at the bottom surface is:
\begin{equation} 
\dot{h}_b=\dfrac{\rho_{sat}\,c_{sat}\,D^{2/3}}{\rho_s\,(\rho_{sat}-c_{sat})} \left(\dfrac{g \cos \theta \, (\rho_{sat}-\rho_f)}{Ra_c\,\nu_i\,\rho_f}  \right)^{1/3}.
\label{eq:hdot} 
\end{equation} 
We evaluate $\dot{h}_b$ using the parameters corresponding to sucrose dissolving in fresh water, where  $\rho_s= 1430$\,g\,L$^{-1}$, $\rho_f =997$\,g\,L$^{-1}$, $\rho_{sat}=1300$\,g\,L$^{-1}$, $c_{sat} = 940$\,g\,L$^{-1}$, $D=4.3 \times 10^{-10}$\,{m$^2$\,s$^{-1}$}, and $\nu_i = \SI{2.5e-4}{\square\meter\per\second}$ by interpolating between the saturation concentration $\nu_s=\SI{7.7e-4}{\square\meter\per\second}$ and the fresh water $\nu=\SI{1.0e-6}{\square\meter\per\second}$~\cite{Wykes2018,Sharma2021}. A comparison is plotted in Fig.~\ref{fig:erosionrate}, and good agreement is observed. 

At the top surface of the slab, the convection is gravitationally stable, and one can obtain (see SI Section~7.~Surface Recession Rates) an expression for dissolution rate averaged over the length of the dissolving surface as:
\begin{equation}
\dot{h}_t=\frac{4B\sin\theta^{1/4}}{3 L^{1/4}},
\label{eq:hdottop}
\end{equation}
where $B$ is a material dependent fitting parameter. This expression is also plotted in Fig.~\ref{fig:erosionrate}, and observed to describe the data reasonably, except near $\theta = 0^\circ$, because the expression is valid only for sufficiently inclined surfaces that give rise to a slope-driven buoyancy flow~\cite{Pegler2020}.

Thus, the body dissolving asymmetrically with $\theta$ with density inversion instability leads to faster dissolution at the bottom surface, which in turn leads to faster $U_b$. At $\theta = 45^\circ$, $\dot{h}$ is approximately 2.5~times higher at the bottom surface compared to the top surface. This factor along with the difference in surface area partly explains the nearly 5~times higher speeds recorded when the bottom versus top surfaces alone are allowed to dissolve as noted in Table~\ref{tab:speeds}.  

\newpage
\section{Propulsion mechanism}
\label{sec:propmech}
\begin{figure*}[t]
\centering
\includegraphics[width=1\textwidth]{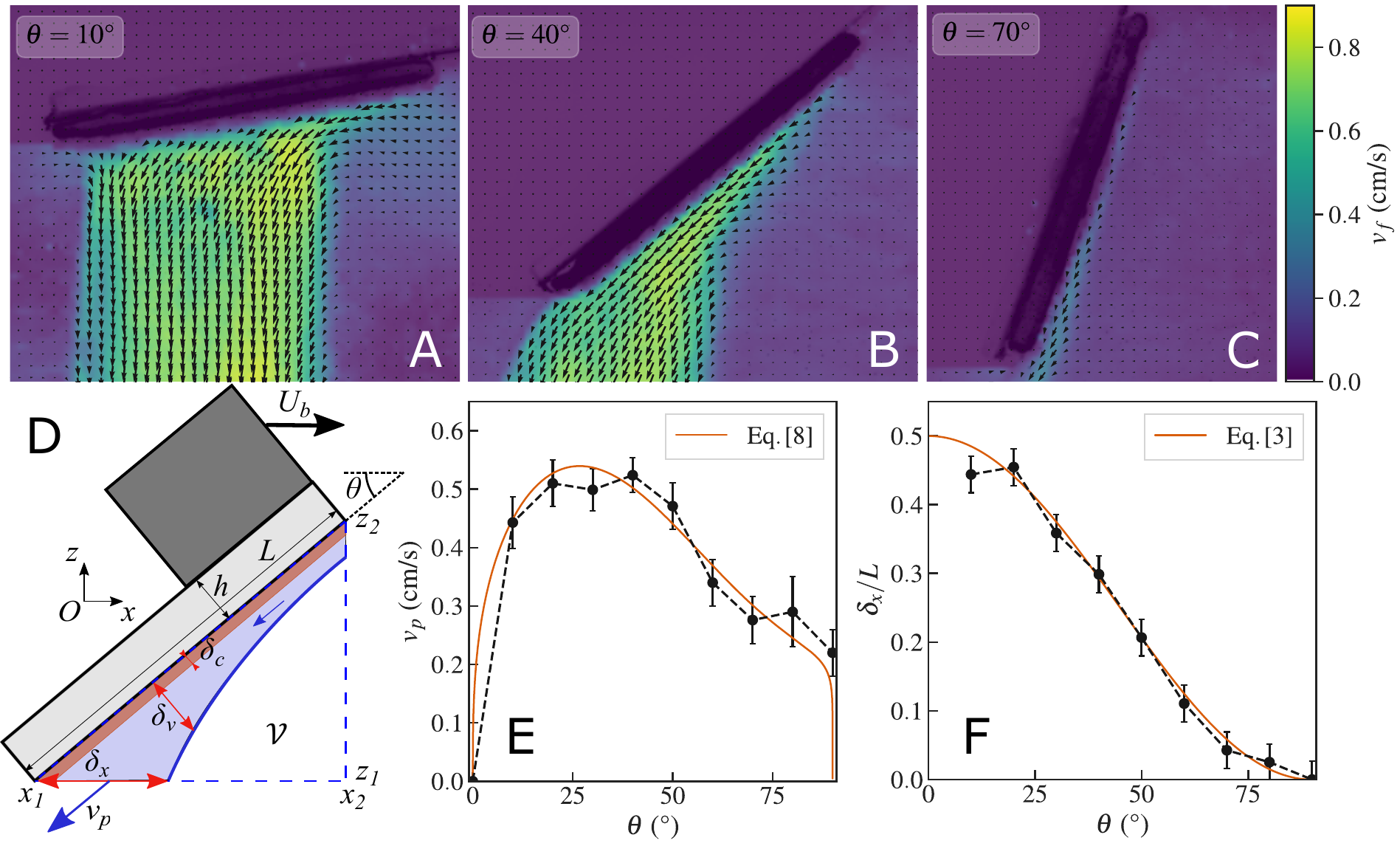}
\caption{\label{fig:piv} Buoyancy driven flow characterization. A-C: Velocity field under the static candy plate at three different inclinations. D:  A schematic of the main features of the flow in the vertical plane important to determining the thrust on the bottom surface of the dissolving slab. A prismatic volume $\mathcal{V}$, denoted by the blue dashed lines with lateral width $W$, below the dissolving surface is used in analyzing the momentum balance. The flow leaving $\mathcal{V}$ through the bottom side is mostly confined over a length $\delta_x$  and has a magnitude $v_p$. The rapid flow is confined over a distance $\delta_v$ perpendicular to the dissolving surface,  and $\delta_c$ is the thin nearly saturated solute boundary layer. E: Plot of $v_p$, corresponding to the maximal value of the parallel component of the fluid velocity at $z = z_1$ between $x_1$ and $x_2$, as a function of $\theta$. Solid orange line corresponds to a theoretical estimation according to Eq.~[\ref{eq:vp}], with a dimensionless prefactor of order one depending on the angle chosen to fit the data ({see SI Section 11. Inverse Friction coefficient and Fig.~S9}). F: The effective width of the rapid flow $\delta_x/L$ versus $\theta$. Solid orange line corresponds to empirical formula given by Eq.~[\ref{eq:deltax}].
}
\end{figure*}

To explain the motion of the dissolving body, we hypothesize that the solute rich density current creates a pressure difference fore and aft of the immersed body giving rise to the thrust needed to accelerate the body from rest (the direct thrust due to the dissolution of mass at a rate given by $\dot{h}$ can be calculated to be negligible as shown in SI Section~8.~Estimate of thrust due to direct solute dissolution). 
When the dissolving surface is located above the solvent, the solute-rich fluid layer is susceptible to a solutal density inversion instability~\cite{Sharma2021}. As seen in the shadowgraph image Fig.~\ref{fig:image}B, the dense fluid does not sink strictly vertically. The gravity-driven current acquires a horizontal component due to a suction effect. Low pressure is created at the bottom face of the slab due to the sinking plumes which causes the boundary layer to remain attached to the surface, while gravity causes it to flow downwards. Thus, the flow must have a horizontal component, directed backwards  relative to the boat. The forward motion of the boat can then be seen in two ways, which are ultimately equivalent, in terms of force or in terms of momentum: either as the result of the net flow directed backwards, or as the result of the low pressure region located along the candy slab which induces a net horizontal force. While buoyancy rapidly ensures vertical balance, the solid is pulled horizontally by the low-pressure region in the direction opposite to the horizontal component of the sinking fluid, providing the propulsion mechanism. If the dissolving surface is located below the solvent, the solute-rich fluid follows the inclined surface in a barely visible thin layer due to unbalanced gravitational force when $\theta > 0$, and is ejected off to the sides as it falls over the inclined dissolving slab. Such a flow would give rise to reaction forces which would further contribute to the propulsion of the boat. Since the downward facing surface dissolves faster and makes a dominant contribution to the observed speed as discussed in Section~1D, we only focus on the region below the dissolving slab to develop a simplified quantitative understanding of the thrust which accelerates the boat. 

A direct evaluation of the pressure field on the solid wall is difficult because of %the intermittency of the 
the intermittent nature of the plumes and the complex flow geometry.  
Therefore, we employ an approach using momentum balance to calculate the thrust from the measured mean velocity field near the dissolving solid. When the boat moves with constant velocity it is not possible to evaluate thrust using the flow field because the thrust equals drag in that limit. However, the thrust can be evaluated from the flow field around a stationary boat, since the thrust remains independent of speed according to Fig.~\ref{fig:Force}A. Therefore, we perform a series of measurements of the flow field below an identical dissolving slab held at rest with various $\theta$. 

Fig.~\ref{fig:piv}A-C shows examples of observed flow fields at low, intermediate, and high inclination angles, respectively, obtained with PIV. We observe that the overall flow is more or less downward along gravity at small $\theta$, but becomes increasingly attached to the dissolving surface with increasing $\theta$. The asymmetry of the slab thus leads the flow to acquire a strong lateral component: this, in addition to being an interesting discovery in itself, is crucial for propulsion. The main features of the solute-rich fluid flow are schematically represented in Fig.~\ref{fig:piv}D, where a steady flow with velocity $v_p$ is directed along the dissolving surface. We denote $\delta_v$ the thickness over which the flow velocity decreases to half its value.  We estimate $v_p$ from the measurements through the horizontal surface below the dissolving body bounded by the $x_1$ and $x_2$ planes, and plot it in Fig.~\ref{fig:piv}E as a function of $\theta$. We observe that $v_p$ increases rapidly as the inclination of the dissolving slab is increased from $\theta = 0^\circ$, and reaches a peak over a similar range as $U_b$, before decreasing as the dissolving surface becomes vertical. Further, we plot the projection $\delta_x$ of the rapidly flowing layer $\delta_v$ on the axis between $x_1$ and $x_2$ scaled by $L$ in Fig.~\ref{fig:piv}F. $\delta_x$ decreases monotonically with $\theta$, and is  well described by the empirical form:
\begin{equation}
   \frac{\delta_x}{L} = \frac{1}{2}\cos^2\theta.
  % \delta_x=(1/2)\,L\,\cos^2\theta
    \label{eq:deltax}
\end{equation} 
It can be noted that $\delta_x=L/2$ when $\theta=0^\circ$, meaning that the width of the rapid current below a horizontal slab is only half that of the slab. Then, $\delta_x/L$ decreases when $\theta$ increases since the flow becomes increasingly narrow and attached to the slab.

Hence, we consider the prismatic volume $\mathcal{V}$, below the inclined dissolving surface, as illustrated in Fig.~\ref{fig:piv}D, bounded by the dissolving surface, the horizontal plane $z=z_1$, and the vertical planes $x=x_2$, $y=y_1$ and $y=y_2$ with $y_{1,2} = \pm W/2$. Assuming the fluid to be inviscid and incompressible, we obtain the horizontal component of the thrust acting on the dissolving surface  considering momentum balance~\cite{GuyonHulinPetit} along the horizontal $x$-axis as 
\begin{eqnarray}
        F_p^{MB} & =  {-} \int_{y_1}^{y_2}\,\int_{z_1}^{z_2}  \rho\, v_x^2  \,\mathrm{d} y \, \mathrm{d} z \,|_{x={x_2}} \nonumber \\ 
 &   + \int_{z_1}^{z_2} \int_{x_1}^{x_2}  \rho\, v_y\,v_x  \, \mathrm{d} x \, \mathrm{d} z \, |_{y=y_1} \nonumber \\
 & - \int_{z_1}^{z_2} \int_{x_1}^{x_2}  \rho\, v_y\,v_x  \, \mathrm{d} x \, \mathrm{d} z \, |_{y=y_2}\, \nonumber \\
&     {+} \int_{y_1}^{y_2}\,\int_{x_1}^{x_2}  \rho\, v_z\,v_x \, \mathrm{d} y \, \mathrm{d} x  \,|_{z=z_1}\, ,
  \label{eq:Fp}
\end{eqnarray}
where $\rho$ is the fluid density and $v_x$, $v_y$, and $v_z$ are the respective velocity components along the $x$, $y$ and $z$ axes as defined in Fig.~\ref{fig:piv}D. The terms on the right hand side correspond to the fluid entering $\mathcal{V}$ through vertical plane $x = x_2$,  bounding vertical planes $y = \pm {W/2}$, and leaving through the bottom horizontal plane $z = z_1$.  Because the mass of dissolved solute is negligible compared to the mass of the fluid, we evaluate each term separately assuming that $\rho$ is given by the density of fresh water $\rho_f$ in $\mathcal{V}$ (see SI Section~9.~Estimate of thrust from momentum balance). We find that the term corresponding to bottom plane dominates over the entire range of $\theta$ (see SI Section 10. Relative thrust contributions and Fig.~S8). Thus, $F_p$ is positive, in agreement with the direction of the acceleration when the body is released. 

We compare the calculated magnitude of $F_p^{MB}$ using Eq.~[\ref{eq:Fp}] with data as a function of $\theta$ in Fig.~\ref{fig:Force}B, and we find that it is overall in agreement with the thrust obtained from both the kinematic as well as the deflection measurements. Thus, we conclude that momentum balance gives a reasonable description of the observed thrust needed to accelerate the body from rest as seen in our experiments. While $F^{MB}_p$ underestimates observed $F_p$ obtained from the kinematics, this is to be expected since we neglect the contribution of the top surface to propulsion, which becomes relatively more important as $\theta$ increases.  

We can further compare the predicted speeds according to momentum balance to those measured in the experiments using the fact that we have $F_p = F_d$, and $F_d = \frac{1}{2}  C_d \rho_f A U_b^2$,
when the boat moves with constant velocity. Then, 
\begin{equation}
  U_b^{MB} = \sqrt{\frac{2 F_p^{MB}}{C_d  \, \rho_f \, A}}\, . 
\label{eq:Ubthe}
\end{equation}
Figure~\ref{fig:image}E shows $U_b^{MB}$ (blue line) compared with measured $U_b$ corresponding to each trial with the 9 different boats. Good agreement is observed with $U_b$, except for $\theta>\SI{60}{\degree}$, where contribution of the attached convection flow on the top of the dissolving slab may become significant. Thus, we find that the mechanism of propulsion based on the pressure variation produced by the attached sinking density current can capture the overall order of magnitude of the thrust acting on the dissolving body, and the speeds attained as a function of the inclination angle.  

\section{Functional analysis of boat speed}
\label{sec:funcUb}
In order to understand the physical and geometrical parameters setting the boat velocity $U_b$ beyond the propulsion mechanism, we further simplify the flow modeling. We consider only the term corresponding to the flow through the bottom surface $z = z_1$ in Eq.~[\ref{eq:Fp}], since it makes the dominant contribution to determine $F_p$ (see SI Fig.~S8). Further, measurements taken in the $y-z$ plane (see SI Section 1. PIV measurements) show that the descending density flow is mainly confined to the central half -width of the dissolving surface. The flow out of $z_1$ is thus mostly confined over a region with length scale $\delta_x$ and width $W/2$, with an approximate flow velocity $v_x = v_p \cos\theta$ and $v_z = v_p \sin\theta$, and we have
\begin{align} 
F_{p}^{MB} \approx \rho_f \frac{W}{2}\,\delta_{x}\,v_p^2\,\sin \theta \,\cos\theta\,.
\label{Eq:Fpest}
\end{align}
Then, using Eq.~[\ref{eq:Ubthe}] and $A = L_A \, W$ (see section 1.C), we have  
\begin{equation}
  U_b^{th} = \sqrt{\frac{\sin 2\theta \, \delta_{x}}{2 \, C_d  \, L_A}} \,v_p\, . 
    \label{eq:Uvp}
\end{equation}
Thus, we observe the body speed is set by $v_p$ besides other geometric factors related to the size and asymmetry of the floating body.  Using Eq.~[\ref{eq:deltax}] to express $\delta_x$ and Eq.~[S1]  to capture the measured dependency of the projected length $L_A$ (see SI Section~5.~Projected Boat Length), we find $\sin2\theta\,\delta_x/L_A $ has a maximum at $\theta \approx 21^\circ$, which equals $0.58$. Thus, we then find $U_b^{th}$ is of similar order of magnitude as %$0.71\,v_p \approx$ 
$v_p$ at  $\theta \approx 20^\circ$ from Eq.~[\ref{eq:Uvp}] since $C_d\approx 0.6$. This analysis which shows that the boat speed is of similar order of magnitude as the density current speed is consistent with our observations in Fig.~\ref{fig:image}E and Fig.~\ref{fig:piv}E at intermediate $\theta$. 

We can estimate $v_p$ itself by balancing the pressure difference which gives rise to the rapid flow below the dissolving slab, starting from $(x_2,z_2)$ and ending at $(x_1,z_1)$, due to the non-buoyant weight of the solute-rich fluid with the drag as a result of 
inertial friction exerted by the quiescent fluid below and the dissolving surface above. Assuming that the shear occurs over length scale $\delta_v$, we have $\Delta \rho\,g\,L\,\sin \theta = f_D\, \rho_f \,\dfrac{L}{\delta_v}\,  v_p^2$, where $\Delta \rho$ is the increase in density due to the dissolved solute, and $f_D$ is a dimensionless friction factor~\cite{GuyonHulinPetit,Schlichting}. 
Using mass conservation of the solute, we have
$\Delta \rho=\dfrac{\rho_s \, \dot{h}\, L}{\delta_v\,v_p}$,  
where $\rho_s$ is the density of the dissolving solid, and $\dot{h}$ is the recession rate of the dissolving surface given by Eq.~[\ref{eq:hdot}]. Thus, 
\begin{equation}
    v_p = \mu_p \left( \frac{\rho_s \,g \, L \, \sin\theta \,\Dot{h}}{  \,\rho_f}  \right)^{1/3},
    \label{eq:vp}
\end{equation}
where $\mu_p = 1/f_D^{1/3}$. Plugging in the values corresponding to where the maximum speed is observed ($\theta =22^\circ$) in $v_p^{o}=\left( \frac{\rho_s \,g \, L \, \sin\theta \,\Dot{h}}{  \,\rho_f}  \right)^{1/3}$, we find that it is $10.1\,$mm/s, comparable in magnitude to the measured $v_p = 5.1\,$mm/s. Thus, $\mu_p$ is a parameter which is of order 1. Because the geometry of the flow evolves with $\theta$, $\mu_p$  is not constant. Thus, we plot the ratio $v_p/v_p^{o}$ in Fig.~S9, and find its dependence on $\theta$. We plot Eq.~[\ref{eq:vp}] with the empirical value of $\mu_p$ as a function of $\theta$ in Fig.~\ref{fig:piv}E. Hence we can estimate the value of $U_b^{th}$ using  Eq.~[\ref{eq:vp}], and the empirical forms for $\delta_x/L$ and $L_A$ as a function of $\theta$. We plot $U_b^{th}$ versus $\theta$ in Fig.~\ref{fig:image}E (orange line), and observe that the estimated values follow the same trends as the data especially over low and intermediate inclination angles. $U_b^{th}$ can be observed to be systematically increasingly lower with increasing $\theta$. This trend is consistent with the expectation that the flow near the top surface becomes increasingly important as $\theta$ increases toward 90$^\circ$, although other assumptions in our model may contribute to these systematic deviations as well. 

\begin{figure*}[t]
\centering
\includegraphics[width=18cm]{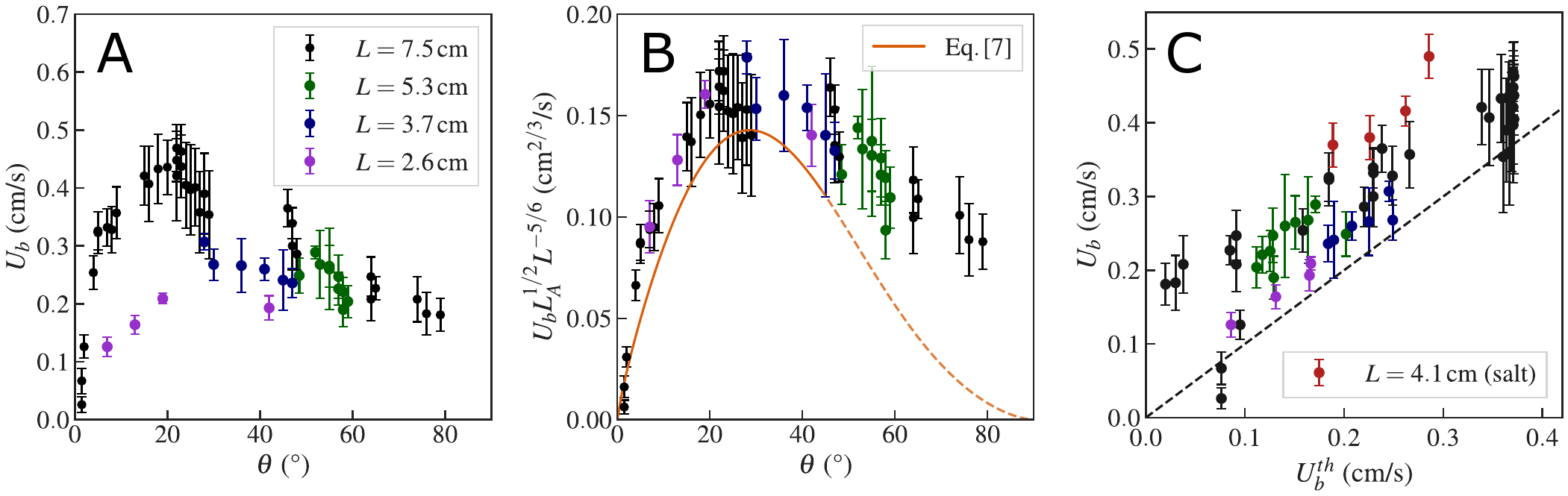}
\caption{\label{fig:SpeedModel} Boat velocity for various sizes and comparison with model. A. Measured $U_b$ corresponding to candy boats with various sizes as a function of $\theta$. B. The data from the various boats collapse after scaling with $L$ and $L_A$. Eq.~[\ref{eq:Uvp}] (solid orange line) matches the data at a 10\% level for $\theta< 45^\circ$, but underestimates the data for higher $\theta$  (dashed orange line) because contributions of flows above the dissolving slab are not taken into account. C. Comparison of measured $U_b$ and $U_b^{th}$ (Eq.~\ref{eq:Uvp}) corresponding to various candy and salt boats of various sizes. The trends are in overall agreement, with measured values being systematically higher compared to calculated values which only consider the contributions of flow below the boat in calculating the thrust. 
}
\end{figure*}

To test the robustness of the measurements and derived dependence further, we built additional boats with $L$ between 7.5\,cm and 2.6\,cm. The data corresponding to each measured trail with the candy boats with various $L$ is plotted in Fig.~\ref{fig:SpeedModel}A versus the measured $\theta$. Comparing $U_b$ at similar $\theta$, we note that speeds are lower for smaller $L$. To understand if this is consistent with our analysis, we combine Eq.~[\ref{eq:deltax}], Eq.~[\ref{eq:Uvp}], Eq.~[\ref{eq:vp}], and the general expression of the erosion rate to obtain the following scaling after neglecting $\theta$-dependence,  
\begin{equation}
    U_b^{th} \approx   \, \left(D\,g\right)^{1/3} \, \left(\dfrac{c_{sat}\,\rho_{sat}}{\rho_f\,(\rho_{sat}-c_{sat})}\right)^{1/3} \delta_c^{-1/3} L_A^{-1/2}\,L^{5/6}.
    \label{eq:Ubth3}
\end{equation} 
The first term constitutes the intrinsic velocity scale of the system, and interestingly gives the right order of magnitude for the boat speed, i.e. $(D \,g)^{1/3}$ is of the order of $1.6\,$mm/s for sugar in water, and $2.5\,$mm/s for salt. Then, we can expect the explicit length dependence to scale as $L^{5/6} L_A^{-1/2}$, which reduces to $L^{1/3}$ when $L_A \approx L$ (see SI Section 12. Calculation of Boat Speed). Accordingly, we plot $U_b L^{-5/6} L_A^{1/2}$ versus $\theta$ in Fig.~\ref{fig:SpeedModel}B, and find that the data collapse onto a common curve, showing that the length dependence is captured by our analysis. Here again, as in Fig.~\ref{fig:image}E, we observe that the theoretical curve given by Eq.~[\ref{eq:Uvp}] is systematically lower compared to the data by about $10\%$ for $\theta < 45^\circ$, but the deviations become larger for higher inclinations with increasing $\theta > 45^\circ$ because only the contribution of the flow below the dissolving slab is considered in the model, and the contribution of the top surface increases.    

In order to compare the measured and calculated speeds explicitly in the case of the various candy boats, but also  the salt boat, we plot $U_b$  versus $U_b^{th}$ in Fig.~\ref{fig:SpeedModel}C after substituting in Eq.~[\ref{eq:vp}] 
with the relevant physical, chemical and geometrical  parameters. Except for the points corresponding to higher $\theta$ which lead to substantially lower $U_b^{th}$, we find that the data collapse close to the unit slope line. Thus, our model of the propulsion mechanism based on pressure difference caused by the rapidly descending density flow is consistent with our observations considering the approximations used to develop the analysis.

\section{Discussions}
In summary, we have demonstrated that an asymmetrically dissolving body can move rectilinearly with significant speeds while floating in a fluid.  Robust directed motion is observed opposite to the principle direction in which boundary layer density currents are set up. We show that propulsion arises because of the differences in pressure fore and aft of the body due to the dense sinking current. The resulting horizontal unbalanced thrust accelerates the body till a constant velocity is reached when the thrust matches the body drag. The observed speed increases initially with inclination angle of the body, which leads to robust generation of gravity current due to break symmetry. The speed then decreases as the horizontal projection of the gravity current decreases with increasing angle leading to an optimal angle when the fastest speeds are observed. This observation of an efficient directed translation motion is in fact not obvious considering the relatively 
small mass loss rate and the intermittence of the flow. We also note that until this study, the convection flows generated by dissolution have not been quantitatively characterized and especially the presence of a horizontal net flow in the asymmetrical case has not been pointed out.

Even though the boundary layer is unstable and the density current is turbulent in our dissolution driven system, the underlying propulsion mechanism at work is similar to that observed in inhomogenously heated bodies that undergo translational motion due to stable laminar buoyancy currents~\cite{Mercier2014}. Thus, the propulsion mechanism is observed across different systems and can be applied to far stronger density currents which result in significantly faster rectilinear speeds. In particular, we find an order of magnitude higher speeds while considering similar centimeter-scale floating bodies, with even greater speeds estimated with increasing size. {Considering observations with boat pairs, %inducing hydrodynamic interactions, such effects %may arise from the interactions between them. For floating bodies, the capillary interactions~\cite{Vella2005,Dalbe2011} may be dominant and 
one can anticipate that capillary and hydrodynamic interactions can lead to collective phenomena such as self-assembly and swarming in dissolving boat clusters. %In addition, a boat should also be affected by the water flows associated to the other boats, inducing hydrodynamic interactions.
}

Because shape evolution due to melting and dissolution can be similar in solids~\cite{Meakin2010,Cohen2016}, it is reasonable to consider whether discernible rectilinear transport can occur during melting or freezing of floating ice blocks due to temperature or salinity gradients~\cite{KerrJFM1994b,KerrJFM1994}. It has been shown that the melting of ice shelves generates upward buoyancy driving sub-glacial plumes before they break into icebergs~\cite{Hewitt2020}. Similar plumes also occur at the immersed edges of icebergs. Recent models~\cite{mcconnochie_kerr_2016} estimate the plume thickness and their velocity to be about 10\,m and 0.6\,m/s, respectively, for an iceberg of submerged height and width of 200 m, under typical conditions observed in the polar regions of the Earth. Similar orders of magnitudes are also obtained for plumes under ice shelves with a typical melting rate of 50 m per year~\cite{Hewitt2020}. %Because $\delta_x$ and $L_A$ scale as the boat length $L$, we expect $U_b^{th}$ to scale with $v_p$ according to Eq.~[\ref{eq:Uvp}]. 

While icebergs come in wide range of sizes, their shapes below the waterline are not known in much detail~\cite{mckenna2005,cenedese2023}. Except for tabular icebergs, the emerged part itself is rarely symmetric due to processes which lead to their formation and deterioration~\cite{romanov2012}. Recent experimental works on the melting of vertical ice columns~\cite{Weady2022} have shown that vertical walls become inclined or scalloped in relation to buoyancy-driven flows with a slope sign depending on the water bath temperature. Further, we have shown that even modest inclination leads to symmetry breaking of the boundary layer flow and directed motion. Therefore, using the corresponding values in the case of ice in place of $v_p$ and $\delta_x\sin \theta$ in Eq.~[\ref{Eq:Fpest}], we find $F_p \approx 7.0 \times 10^3$\,N.  By comparison, a wind of moderate speed $U_W \sim 5$ m/s exerts on an exposed surface of area $S_{a}=10\times 200$ m$^2$, a force $F_W \approx \rho_{air}\,S_{a}\,U_W^2 \approx 5\times 10^4$\,N. {This would imply a corresponding additional iceberg drift speed of approximately 2\,cm/s. Thus a typical 200\,m sized iceberg would move over its body length within about 3\,hours. Such an appreciable effect should be measurable in the field by placing a couple of Global Positioning System (GPS) devices to track the location and rotation rate of a set of icebergs, while measuring the local wind and marine currents. The subsurface iceberg shape can be obtained with Ground-Penetrating Radar (GPR)~\cite{Bohleber2017}, or with sonar placed on autonomous submarines~\cite{Zhou2019}. Then, by subtracting off the contributions of wind, currents, waves, and Coriolis force~\cite{Mountain1980,Anderson2016,Wagner2017,Marchenko2019}, it may be possible to ascertain the importance of gravity-current driven propulsion by correlating them to rapidly melting icebergs with a statistically significant number of measurements. This would establish that contribution of ice melting on its drift may be not negligible, for a typical asymmetrical iceberg with inclined immersed walls.} Thus, while wind and currents can lead to significant transport of sea ice and icebergs, density currents due to salinity gradients generated by iceberg melting, may in principle contribute to their transport as well.

\section*{Methods}

\subsection{Boat Construction}
The boats are assembled by attaching a dissolving slab (dimensions $75\,{\rm mm} \times 40\,{\rm mm} \times 5\,{\rm mm}$, mass $\approx\SI{25}{\gram}$) to a hollow plastic box (dimensions $30\,{\rm mm} \times 40\,{\rm mm} \times 35\,{\rm mm}$ with $\SI{1.5}{\milli\meter}$ thick walls, mass $\SI{10}{\gram}$) with a thin layer of silicone sealant. Candy slabs are prepared following protocols used in previous studies~\cite{Huang2020,Sharma2021} to cast an inexpensive, homogeneous, reproducible and fast dissolving material with a prescribed shape.  Rectangular shaped slabs are prepared by blending granulated sugar, light corn syrup and water starting with a $8:3:2$ volume ratio. The mixture is then heated up to $150^\circ$C and poured into silicone molds with rectangular cross sections, which after cooling result in solid slabs with requisite dimensions. The plastic box is 3D-printed with PolyEthylene Terephthalate (PET) filament with a Prusa 3D MK3S printer. 
Additional experiments are performed with smaller slabs with dimensions $53\,{\rm mm} \times 27\,{\rm mm} \times 5\,{\rm mm}$, $37\,{\rm mm} \times 20\,{\rm mm} \times 5\,{\rm mm}$ and $26\,{\rm mm} \times 14\,{\rm mm} \times 5\,{\rm mm}$.  
The salt (NaCl) slab with dimensions $41\,{\rm mm} \times 23\,{\rm mm} \times 6\,{\rm mm}$ is an optical rectangle window for Infrared Spectroscopy obtained from Alfa Aesar.\\

\subsection{Experimental Protocols}
Experiments to measure speeds are performed in a rectangular $\SI{75}{\centi\meter} \times \SI{45}{\centi\meter}$ tank filled with distilled water to a depth of $\SI{20}{\centi\meter}$. Initially, a boat is gently placed by hand at one end of the tank. Its motion is then recorded by a camera located above the tank, with a frame rate of 5 fps. After $\SI{180}{\second}$, the boat is taken out of the water. The boat weighs typically $\SI{3}{\gram}$ less after each launch, which corresponds to the amount of mass dissolved during the experiment. Measurements are performed between 3 and 5 times for each boat at which point the decrease of the size of the candy plate ceases to be negligible, and the dissolving slab is replaced.

\begin{acknowledgments}
We thank Sylvain Courrech du Pont (Universit\'e Paris Cit\'e), Olivier Devauchelle (IPG Paris) and Ramiro Godoy-Diana (ESPCI Paris) for scientific discussions, as well as Animesh Biswas (Clark University) for help in performing these experiments. This research was partially funded by the ANR grants Erodiss ANR-16-CE30-0005 and PhysErosion ANR-22-CE30-0017, and by U.S. NSF Grant No. CBET-1805398. M.C. and M.B. acknowledge the hospitality of the Department of Physics at Clark University while the initial experiments were developed and performed.
\end{acknowledgments}

% Bibliography

\clearpage

\onecolumngrid

\appendix
\hspace{2cm}
\begin{center}
 \Large{\textbf{Supplemental material}}
\end{center}

\section{PIV measurements}
\label{sec:piv} 

To visualize the fluid flow, the bath near the boat was seeded with neutrally buoyant micron-sized polystyrene tracer particles which fluoresce when illuminated with 532\,nm green light. A cylindrical lens attached to a laser is used to illuminate vertical planes near the dissolving surface of the boat, parallel and perpendicular to direction of motion. A Phantom VEO-E\,310L\,camera is then used to obtain a set of images with a prescribed frame rate and time duration. 
Then, the velocity field of the fluid is obtained by Particle Image Velocimetry (PIV) using the OpenPIV Python package~\cite{OpenPIV} in a grid spacing of $\SI{1.4}{\milli\meter} \times \SI{1.4}{\milli\meter}$.

\begin{figure}
\centering
\includegraphics[width=0.66\columnwidth]{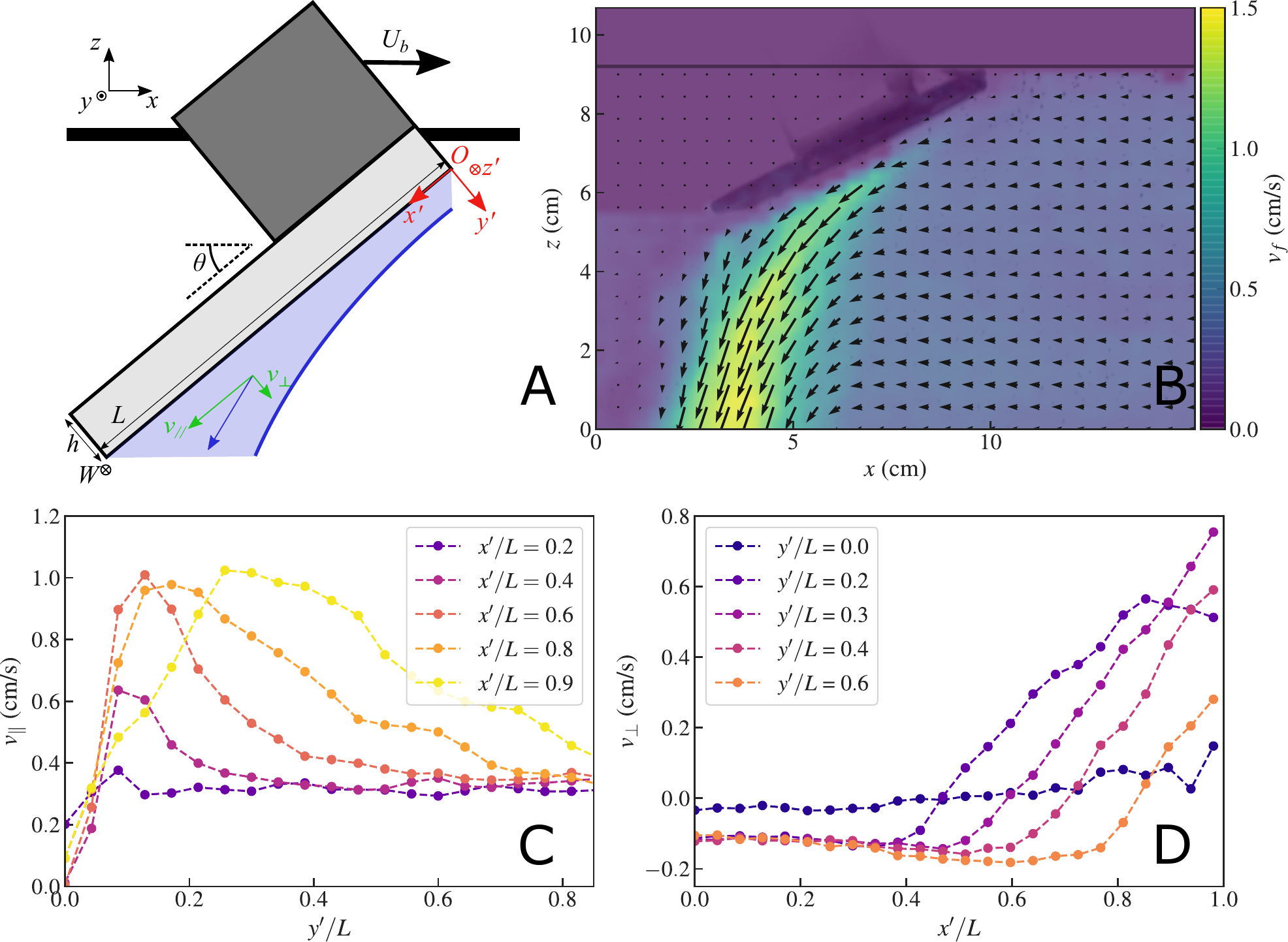}
\caption{\label{fig:velprofile} A. Schematics of the system used to perform PIV measurements, and the coordinate system used to plot the velocity as a function of distance parallel and perpendicular to the dissolving surface. B. An example of a velocity field corresponding to $\theta = \SI{26}{\degree}$ while the boat moves with speed $U_b = \SI{0.4}{\centi\meter\per\second}$. C. The velocity parallel to the bottom dissolving surface $v_\|$ as a function of $y'$ at various distances $x'$ from the bow. D. The velocity perpendicular to the bottom dissolving surface $v_\perp$ as a function of $x'$, at various distances $y'$. This graph can be used to infer the detachment length of the boundary flow $L_p$  In this case, $L_p \simeq \SI{3}{\centi\meter}$. 
}
\end{figure}

An example of the velocity field of the fluid around the moving boat corresponding to $\theta=\SI{26}{\degree}$ is shown in Fig.~\ref{fig:velprofile}A. Here, the images were taken over a 15\,s time interval with a frame rate of 50\,Hz, and then cropped to center the moving boat in the frame. The coordinate system $(x', y', z')$  in the body frame of reference is also shown. 
Its origin is located at the highest central point of the bottom surface. The inclination of the boat $\theta$ is the angle between $O_{x'}$ and the horizontal, or equivalently, the angle between $O_{y'}$ and the vertical. Thus, we can define two components of the velocity field, $v_{\mathit{\parallel}}$ parallel to $O_{x'}$ and $v_\perp$ parallel to $O_{y'}$.
Fig.~\ref{fig:velprofile}B and Fig.~\ref{fig:velprofile}C show corresponding plots of the velocity components $v_{\|}$ and $v_{\perp}$, respectively parallel and perpendicular to the dissolving surface. These plots show that the flow initially follows the surface of the slab, remaining essentially parallel to it, before detaching from the surface as $x'/L$ increases above $0.5$. We note that  $v_{\|}$ increases in strength rapidly with $x'/L$, before reaching a peak which is about 2.5 times the boat speed $U_b$. It can be further noted that $v_{\|}$ and $v_{\perp}$ increase away from the surface as the flow detaches.

A set of measurements of the flow were also performed near the dissolving candy slabs with dimensions 7.5\,cm $\times$ 4\,cm $\times$ 0.5\,cm, while they were held in a water bath at various angles between 0 and $\SI{90}{\degree}$. In this case, the flow was visualized at a frame rate of $\SI{100}{\hertz}$ using a Phantom V1840 camera and $\SI{50}{\micro\meter}$ polyamide particles, and then averaged over 50\,s (5000 frames) to obtain the mean flow below the dissolving slab. Figs.~4ABC in the main text show the 2D vertical velocity field in the $xz$-plane corresponding to low, intermediate and high inclination angles. The velocity parallel to the bottom dissolving surface of the boat $v_p$ and the associated width $\delta_x$ obtained using these measurements are plotted as a function of $\theta$ in Figs.~4EF in the main text.

\begin{figure}
\centering
\includegraphics[width=0.66\columnwidth]{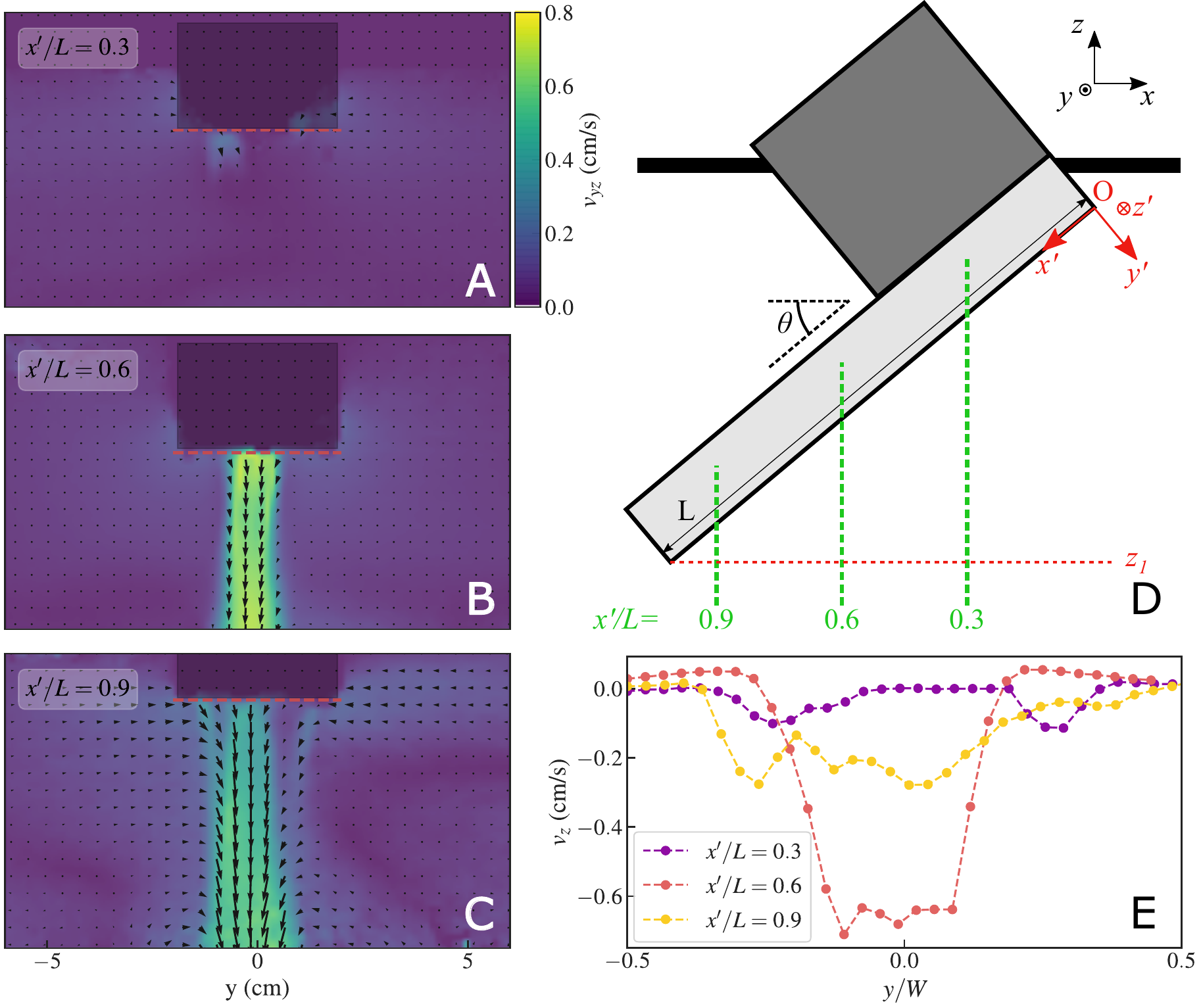}
\caption{\label{fig:piv_candyplate1} A-C. The fluid velocity field under a candy slab (front view) inclined at an angle $\theta=20^\circ$. Vertical planes located at the distance {$x'/L =0.3, 0.6, 0.9$} from the front of the boat are illuminated by the laser sheet in these examples. D. Schematic of the system, with coordinates system and location of the three planes on which PIV is performed. E. The vertical velocity $v_z$ versus $y$ at three different planes, at $z=z_1$.}
\end{figure}

Fig.~\ref{fig:piv_candyplate1}A-C show examples of the flow field obtained in three $yz$-planes through the slab corresponding to increasing $x'$ from the bow. The vertical component of the fluid velocity $v_z$ is plotted corresponding to the plane $z=z_1$, just below the bottom end of the dissolving surface in Fig.~\ref{fig:piv_candyplate1}D. We observe that the flow is mainly located in a region that is about half the block width $W$ and grows with increasing $x'$. We use these observations to estimate the area over which the thrust acts to calculate the propulsive force (see Section~2 and~3 in the main text).

%\newpage

\hspace{1cm}
 
\clearpage
\section{Fully submerged body}

\begin{figure}[h!]
\centering
\includegraphics[width=.8\columnwidth]{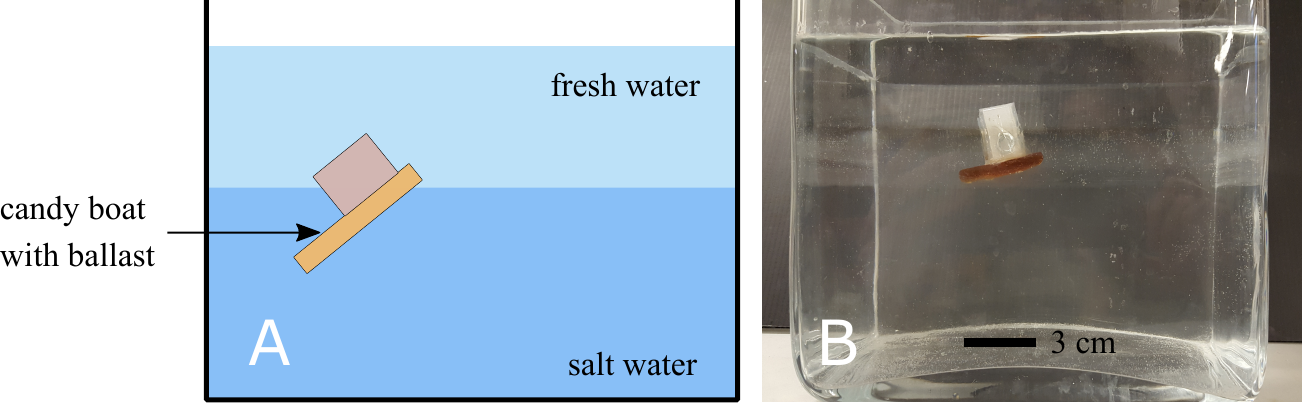}\
\caption{\label{fig:submarine} A: Schematic of the experimental setup used to investigate motion of a fully submerged dissolving body. B: Image of the fully submerged body in the stratified bath. %\red{Add a scale bar if possible.}
}
\end{figure}

In order to verify that the propulsion of the boat is not due to a surface phenomenon (such as the Marangoni effect), we built a candy submarine. To do this, we added ballast inside the hollow buoy of a candy boat ($L = 3.75$\,cm and $W = 2$\,cm) and placed it in a stratified bath. The bath consists of a $\SI{4}{\centi\meter}$ deep layer of fresh water (density $\rho=\SI{1000}{\kilo\gram\per\cubic\meter}$) on top of a $\SI{12}{\centi\meter}$ deep layer of salt water (density $\rho_{sw}=\SI{1100}{\kilo\gram\per\cubic\meter}$). By adding small pieces of metal as ballast, we achieved an  average body density 
between the density of the two fluids. Consequently, the boat floats at the interface between the two stratified layers of fluid. The boat is observed to move rectilinearly with a speed of about $\SI[per-mode=symbol]{0.1}{\centi\meter\per\second}$, which is of the same order of magnitude as velocities observed for floating boats with similar dimensions.

\section{Effect of Dissolving Material}

\begin{figure}[h!]
\centering
\includegraphics[width=.75\columnwidth]{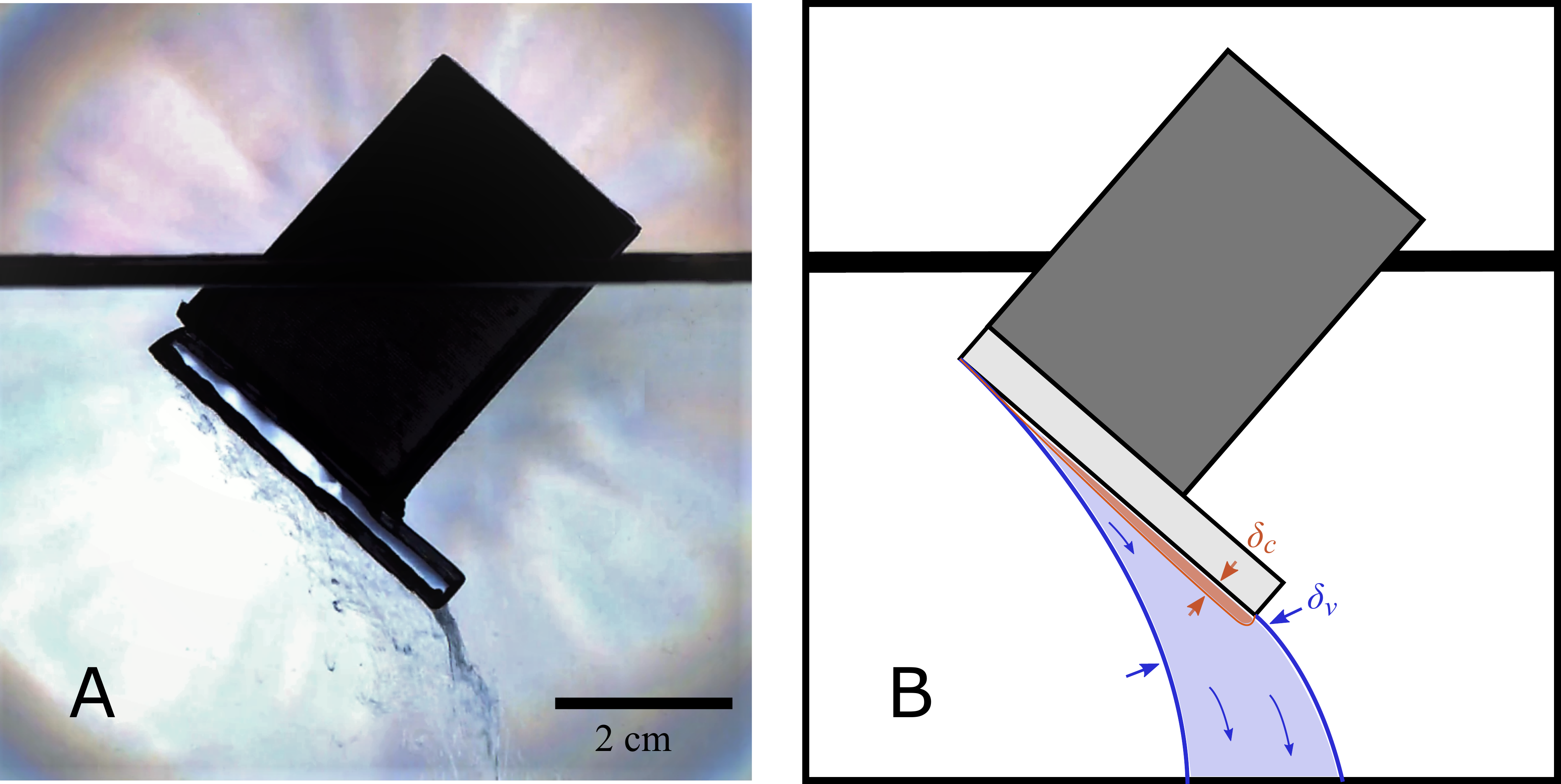}\
\caption{\label{fig:saltboat} A: Image of a salt boat obtained with shadowgraph imaging ($\theta = \SI{43}{\degree}$). B: Simplified schematic of the boat with main features of the flow. The solute concentrated boundary layer of thickness $\delta_c$ destabilizes, which creates a convective flow of typical thickness $\delta_v$, similar to that observed with candy boats.  
}
\end{figure}

We constructed dissolvable boats with salt (NaCl) slabs instead of sugar, with otherwise similar design, to check the effect of dissolving material on the observed dynamics. A crystalline salt slab of density $\rho_s = \SI{2348}{\kilo\gram\per\cubic\meter}$ with dimensions $4.1\,{\rm cm} \times 2.3\,{\rm cm}\times \SI{0.6}{\centi\meter}$, manufactured by Alfa-Aesar (now Thermo Scientific Chemicals), is attached to a $2.0\,{\rm cm}\times 2.7\,{\rm cm} \times \SI{3.5}{\centi\meter}$ hollow 3D printed plastic box. While floating in water, this salt boat has an inclination angle $\theta = \SI{43}{\degree}$. 
In the case of NaCl, the saturation concentration $c_{sat}=317$\,kg\,m$^{-3}$, the density of the saturated solution $\rho_{sat}=1200$\,kg\,m$^{-3}$, the diffusion coefficient $D = 1.61 \times 10^{-9}$\,m$^2$\,s$^{-1}$, and the kinematic viscosity of the saturated solution is $2.0 \times 10^{-6}$\,m$^2$\,s$^{-1}$~\cite{Handbook}.

Figure~\ref{fig:saltboat}A shows an image of the moving salt boat with shadowgraph technique to visualize the density current which arises near the dissolving surface. As in the case of the candy boat, plumes are observed to descend below and behind the dissolving slab. The boundary layer formed near the up-facing surface is too thin to be visible. The boundary layer formed below the dissolving surface and its flow is shown in Fig.~\ref{fig:saltboat}B. We observe a boundary layer thickness $\delta_v \approx 1$\,cm. While the resulting density difference of saturated salt solution is similar to that of saturated sugar solution, the relative viscosity change is only a factor two, in contrast with that of saturated sugar solutions where it changes by at least two orders of magnitude. Nonetheless the boat is observed to move with a speed $U_b = \SI[per-mode=symbol]{0.42}{\centi\meter\per\second}$, similar in magnitude to those observed with a candy boat with similar dimensions and inclination angles.  Thus, we conclude that the propulsion mechanism is independent of the exact nature of the solute and the viscosity of the resulting solution.

\section{Boat Pairs}

\begin{figure}[h!]
    \centering
    \includegraphics[width=0.75\columnwidth]{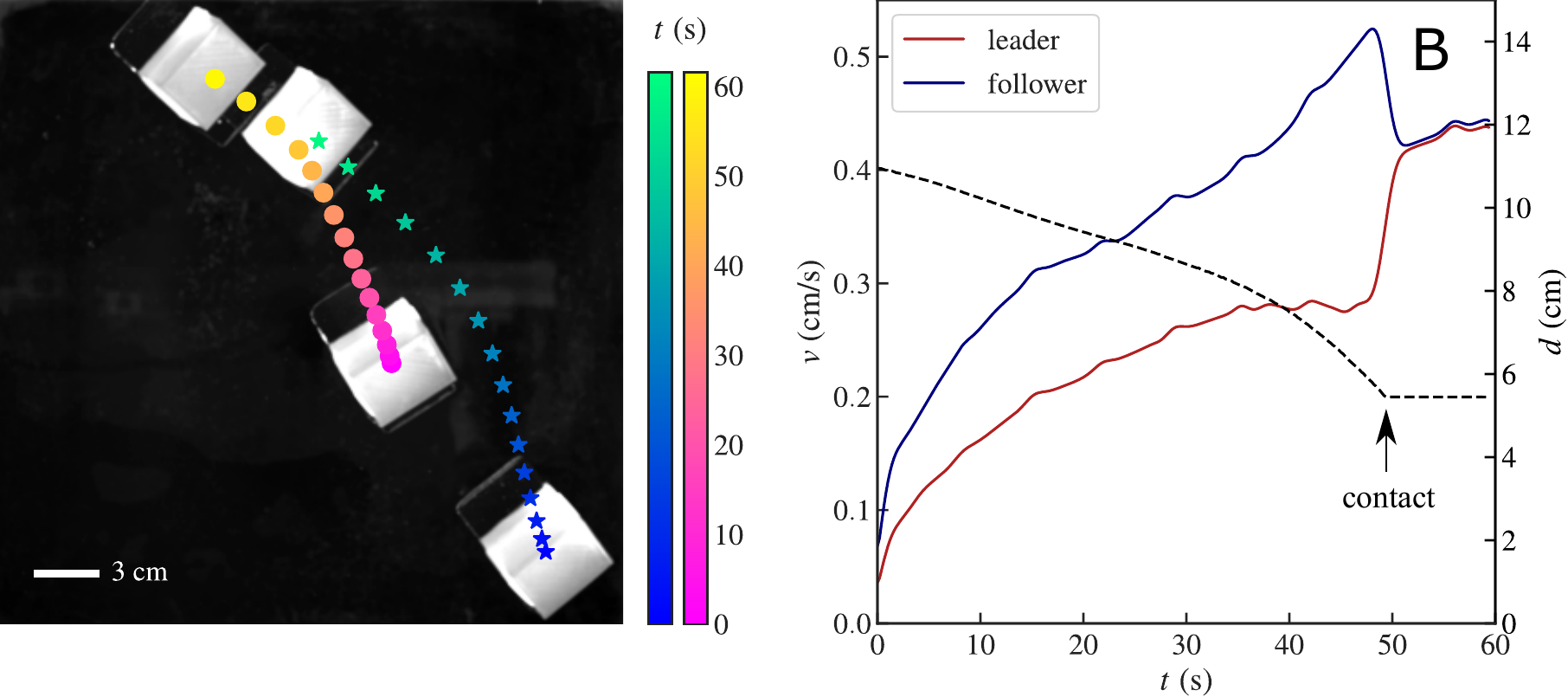}
    \caption{{A: Superposed positions of two identical candy boats while following each other (top view). The following boat is observed to catch up with the leading boat. Dots and stars indicate positions of the center of mass of the leader and follower boats, respectively, at 4\,s time intervals at times indicated by the color bar. B: The distance between the two boats (black dotted line) decreases until they make contact, then remains constant. The velocity of the two boats (red and blue solid lines) first increases as they accelerate from rest and cross the water tank (solid lines). The follower is observed to move faster until it catches up with the leader ($L = 7.5$\,cm; $\theta = 33^o \pm 2^o$).}} 
    \label{fig:pair}
\end{figure}

\begin{figure}
    \centering
    \includegraphics[width=0.75\columnwidth]{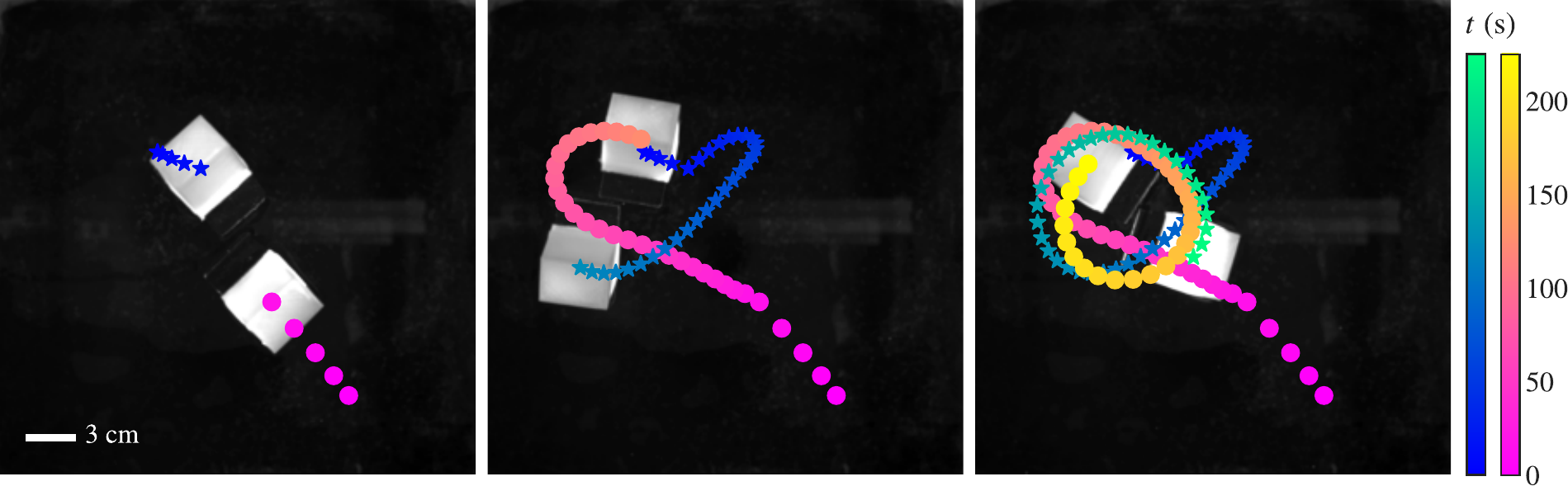}
    \caption{{Position of two boats while moving towards each other from opposite directions (panel 1, top view). The boats are observed to spin around each other after coming in contact (panel 2 and 3). The position of the center of mass of each boat is denoted by circles and stars at 4\,s time intervals  at times indicated by the color bars ($L = 7.5$\,cm; $\theta = 33^o \pm 2^o$).}}
    \label{fig:pair2}
\end{figure}

{We investigated interactions between dissolving bodies by performing experiments with two identical floating candy boats with the same length $L = 7.5$\,cm, and inclination $\theta = 33^o \pm 2^o$ which typically move with a speed of 0.4\,cm/s. The experiments were performed in a $40 \times \SI{40}{\centi\meter}$ tank filled with water to a height of $\SI{30}{\centi\meter}$. The boats were released from rest at the same time. Depending on the initial relative position and orientation of the boats, a variety of outcomes are observed. 

When the boats are placed one behind the other at a distance of a few centimeters, the following boat accelerates faster than the leader and eventually catches up with it. The floating boats are attracted by capillary interactions~\cite{Vella2005,Dalbe2011} and remain in contact, even if the contact point is below the free surface. The self-assembled cluster of two boats then moves forward in tandem (see Fig.~\ref{fig:pair} and Movie~S5) with a velocity of about $0.45$\,cm/s. If the boats are placed approximately opposite each other, they move closer together until they collide and then stick together due to opposite thrust forces, friction, and possibly capillarity. If their axes are not perfectly aligned, the boats rotate slowly around each other (see Fig.~\ref{fig:pair2} and Movie~S6) with angular velocity of approximately 0.82~revolutions per minute and a tangential speed of 0.65\,cm/s. Even greater angular velocities are observed if the boats attract each other as they pass each other due to capillarity and stick together on their sides. In the example shown in Movie~S7, the boats perform 2.3 revolutions per minute with a tangential speed of 0.96 cm/s. These velocities are comparable, if not greater, than those measured with a single boat with similar $\theta = 33^o$. 

Thus, we find that the dissolution-driven propulsion occurs robustly even in the presence of other boats, with different modes of collective behavior, depending on their initial approach directions.}

%{We investigated possible interactions between boats by performing experiments with two identical candy boats with the same length $L = 7.5$\,cm, and inclination $\theta = 33^o \pm 2^o$. The experiments were done in a $40 \times \SI{40}{\centi\meter}$ tank filled with water to a depth of $\SI{30}{\centi\meter}$. The boats were released from rest at the same time. Depending on the initial relative position and orientation of the boats, an interesting diversity of behavior can be observed. When the boats are placed one behind the other at a reasonable distance, the following boat moves faster than the leader and eventually catches up with it. Then, the boats stay together by capillary attraction and move forward in tandem (see Fig.~\ref{fig:pair} and Movie~S5). If the boats are placed approximately opposite each other, they move closer together until they collide and then stick together. If their axes are not perfectly aligned, the boats rotate slowly around each other (see Fig.~\ref{fig:pair2} and Movie~S6). Even greater angular velocities can be obtained if the boats stick together by their sides (see Movie~S7).}

\newpage

\section{Projected Boat Length 
}
\label{section:LA}
\begin{figure}[h!]
\centering
\includegraphics[width=0.5\columnwidth]{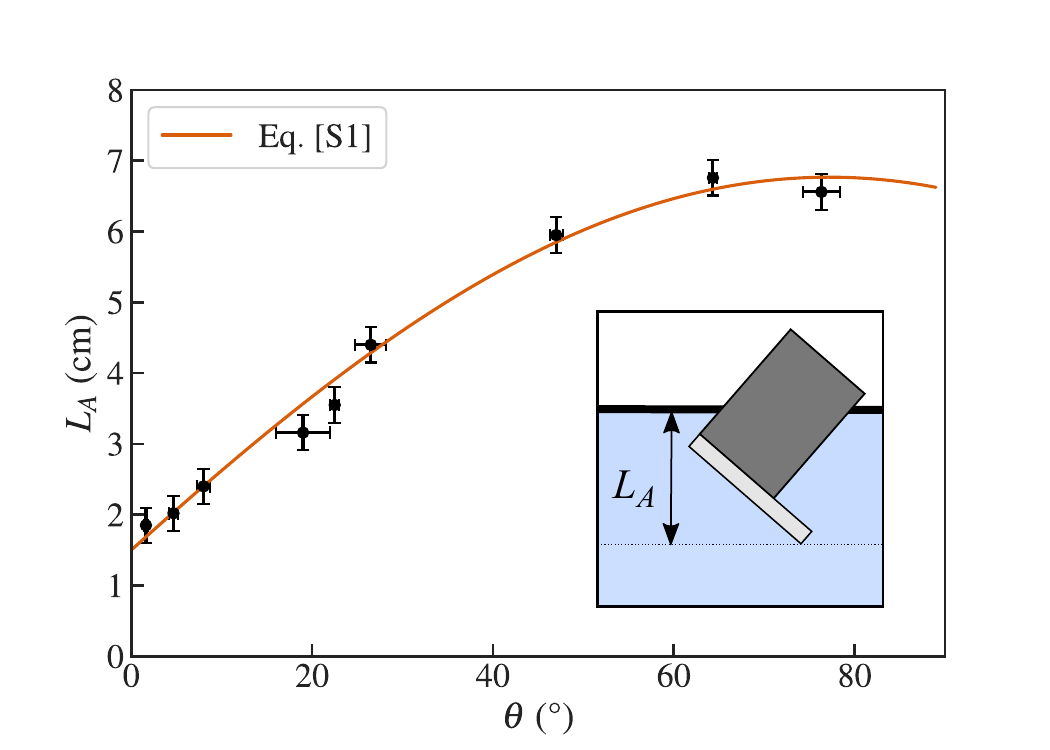}
\caption{\label{fig:proj} The measured projected length $L_A$ ($\bullet$) as a function of the boat inclination $\theta$  and a fit (orange line) given by Eq.~[\ref{eq:proj}]  with $L_1=6.6$\,cm and $L_2=1.5$\,cm. The error bars corresponding to the measurement of $L_A$ and $\theta$ are also shown. Inset: A schematic showing the projected length $L_A$.}
\end{figure}

Figure~\ref{fig:proj} shows the measured projected length $L_A$ of the immersed portion of the body projected on the vertical axis as a function of the inclination angle $\theta$. Each $L_A$ is obtained from side view images of the nine boats while floating near the air-water interface. As expected for a long rectangular shape, $L_A$ increases with $\theta$. We find a linear expansion in terms of sinusoidal functions,
\begin{equation}
L_A=L_1\,\sin\theta+L_2\,\cos\theta,
\label{eq:proj}
\end{equation}
with $L_1=6.6\,$cm and $L_2=1.5\,$cm can describe the data fairly well. The sine and cosine terms can be interpreted as arising essentially from the projection of the boat length and immersed portion of the buoy on to the vertical plane, respectively. Then, we obtain the projected area $A = L_A \, W$, where $W$ is the boat width. This is used to estimate the inertial drag of the floating body as discussed in Section~1 in the main text. 

\newpage
\section{Net Force Evolution}
We assume that the boat accelerates as a result of a constant propulsive force $F_p$ and inertial drag which increases quadratically with speed. %We denote its velocity $v$, projected on its symmetry axis. 
Then, force balance gives us
\begin{equation*}
m\frac{\mathrm{d}v}{\mathrm{d}t}=F_p - \frac{1}{2} C_d \rho A v^2,
\end{equation*}
where $m$ is the mass of the boat, $v$ the velocity of the boat at time $t$, $C_d$ the drag coefficient, $\rho$ the density of water, and $A$ the projected area of the boat along the direction of the boat's velocity.
This equation can be analytically solved by separation of variables
\begin{equation*}
\frac{m \mathrm{d}v}{F_p - \frac{1}{2} C_d \rho A v^2}=\mathrm{d}t,
\end{equation*}
%which we rewrite, for the sake of simplicity, as
%\begin{equation*}
%\frac{dv}{P - Q v^2}=dt,
%\end{equation*}
%where $P=F_p/(\gamma m)$ and $Q=\frac{1}{2} C_d \rho A /(\gamma m)$ are positive and constant. Rewriting as a partial fraction gives
%\begin{equation*}
%dt=\frac{dv}{P - Q v^2}=\frac{dv}{2\sqrt{P}}\left( \frac{1}{\sqrt{P}+\sqrt{Q}v} + \frac{1}{\sqrt{P}-\sqrt{Q}v} \right).
%\end{equation*}
%Integrating, we have
%\begin{equation*}
%t-t_0=\frac{1}{2\sqrt{PQ}} \ln \left[ \frac{\sqrt{P}+\sqrt{Q}v}{\sqrt{P}-\sqrt{Q}v} \right].
%\end{equation*}
%Now, assuming the boat is released from rest, $v=0$ at $t_0=0$. Rewriting in terms of $v$ as a function of time, we have
%\begin{equation*}
%v=\sqrt{\frac{P}{Q}}\left[\frac{\exp \left(t / \tau \right)-1}{\exp\left(t / \tau \right)+1}\right],
%\end{equation*}
%where $\tau=\frac{1}{2\sqrt{PQ}}$. Substituting for $P$ and $Q$, we obtain
assuming that the boat starts from rest at time $t=0$, as 
\begin{equation*}
v(t)=\sqrt{\frac{2F_p}{C_d \rho A}}\left[\frac{\exp \left(t / \tau \right)-1}{\exp\left(t / \tau \right)+1}\right],
\end{equation*}
where $\tau=\frac{m}{\sqrt{2 F_p C_d \rho A}}$. Because the net force  $F_{\mathrm{net}} = F_p - \frac{1}{2} C_d \rho A v^2$, we obtain
\begin{equation}
F_{\mathrm{net}}(t)=F_p\left( 1-\left[\frac{\exp \left(t / \tau \right)-1}{\exp\left(t / \tau \right)+1}\right]^2\right).
\label{eq:Fnet}
\end{equation}

This equation is used in Fig. 2A of the main text to fit the measurement of $F_{\mathrm{net}}$ from the kinematic measurements in order to deduce $F_p$.

\section{Surface Recession Rates}
The dissolution rate of a solid surface depends on its orientation relative to gravity, and whether the concentrated boundary layer is attached or detached. 

\subsection{Dissolution rate at the slab bottom surface}
We summarize here the physical reasoning used to calculate the dissolution rate when the concentrated boundary layer is detached due to a density inversion instability following Sharma, {\it et al.}~\cite{Sharma2021}. The recession rate is obtained by writing  conservation laws at the interface in terms of the position of the interface $h$ relative to the initial surface. Mass conservation gives
$$-\rho_s \frac{\mathrm{d}h}{\mathrm{d}t}={\rho_i}\,\left(\mathbf{u_i}\cdot \mathbf{n} -  \frac{\mathrm{d}h}{\mathrm{d}t}\right)\, ,$$ with $\mathbf{u_i}$ the fluid velocity at the interface, $\mathbf{n}$ a unitary vector normal to the interface, $\rho_s$ the density of the solid, and $\rho_i$ the density of the liquid at the interface. Then, the conservation of the solute gives
$$\rho_s \frac{\mathrm{d}h}{\mathrm{d}t}=\left(\frac{\mathrm{d}h}{\mathrm{d}t}-\mathbf{u_i}\cdot \mathbf{n} \right)\, c_i + D\,\mathbf{\nabla} c \cdot \mathbf{n}\, , $$ where $c$ is the mass concentration of solute in the solution, and $c_i$ is the concentration of solute at the interface. The last term corresponds to the diffusive flux at the interface according to Fick's law. For fast dissolving solids such as sugar and salt in water, the interface concentration $c_i$, to a good approximation, is very close to the saturation concentration $c_{sat}$~\cite{Philippi2019}, and thus $\rho_i$ is also close to the saturation density $\rho_{sat}$. Assuming that the bath is solute-free, and $\delta_c$ is the thickness of the concentrated solute boundary layer, the diffusive flux $D\,\mathbf{\nabla} c \cdot \mathbf{n}$ can then be approximated as $D\,c_{sat}/\delta_c$. By combining these two equations, we obtain the dissolution rate along the bottom of the block
\begin{equation}
\dot{h_b} = \frac{\mathrm{d}h}{\mathrm{d}t} = \frac{D \, c_{sat}}{\rho_s \, \delta_c \, (1-c_{sat}/\rho_{sat})}\,.
    \label{eq:rrates}
\end{equation}
%where %the solid density is $\rho_s = 1400$ kg/m$^3$, 
%the saturation concentration of sucrose $c_ {sat} = 940$\,kg/m$^3$, $\rho_s=1410$\,kg/m$^3$, and the liquid density at saturation $\rho_{sat} = 1300$ kg/m$^3$~\cite{Wykes2018}, while starting with distilled water as the solvent. 

\subsection{Dissolution rate at the slab top surface}
The recession rate of a surface with concentrated boundary layer which remains attached to surface was estimated by Pegler, {\it et al.}~\cite{Pegler2020} as
\begin{align*}
\Dot{h}\left(s,t\right)=\frac{B\,(\sin\left[\theta(s,t)\right])^{1/3}}{\left(\int_0^s\,(\sin \left[\theta(s,t)\right])^{1/3}\mathrm{d}s \right)^{1/4}},
\end{align*}
where $\theta$ is the surface inclination angle, $s$ the curvilinear distance from the top edge of the surface, and $B$ is a convective strength which depends on the properties of the fluid and the dissolving solid.  

In case of the dissolving slabs, $\theta$ is observed to be nearly constant (at least in the beginning), and thus the expression simplifies to
\begin{align*}
\Dot{h}\left(s\right)=\frac{B\,(\sin \theta)^{1/4}}{s^{1/4}}.
\end{align*}
Then, the average dissolution rate $\Dot{h}$ over the slab length $L$ is given by
\begin{align}
\Dot{h_t}=\frac{1}{L}\int_0^L\Dot{h}\left(s\right) \mathrm{d}s=\frac{4B\,\left(\sin \theta\right)^{1/4}}{3L^{1/4}}.
\label{Eq:htop}
\end{align}
Eq.~[\ref{Eq:htop}] can be observed to well describe the data shown in Fig.~3 in the main text, and gives $B=1.15 \times 10^{-4}$\,cm$^{5/4}$ s$^{-1}$, close to the value obtained in Ref.~\cite{Pegler2020} for a cone ($B\simeq 1.5 \times 10^{-4}$\,cm$^{5/4}$ s$^{-1}$).

\section{Estimate of thrust due to direct solute dissolution}
We estimate the thrust due to the direct ejection of the dissolving solute corresponding to a propulsion mechanism as in a rocket. This thrust can be estimated using momentum balance along the direction of motion using the rate of loss of mass due to dissolution. 
Considering the bottom surface of the dissolving slab, and its recession rate $\Dot{h}$, the rate of loss of mass $\Dot{m} = \rho_s\,W\,L\,\Dot{h}$, where $\rho_s$ is the density of dissolving medium. %we have 
Then, the upper bound for the estimated thrust can be arrived by assuming that the dissolved medium moves with the maximum recorded flow speed adjacent to the dissolving surface $v_p$. In this case, the reactive thrust 
\begin{equation*}
F_{r}=\rho_s\,L\,W\,\Dot{h} \, v_{p}\,\cos\theta \,.   
\end{equation*}
For the candy slab used in the experiments, $\rho_s=\SI{1430}{kg\per\cubic\meter}$, $L=\SI{7.5}{\centi\meter}$,  $W=\SI{4}{\centi\meter}$, $\theta=\SI{26}{\degree}$, $v_{p} \approx \SI{0.5}{\centi\meter\per\second}$ (see Fig.\,4E in the main text), and $\Dot{h} \approx \SI{2.5}{\micro\meter\per\second}$ (see Fig. 3 of the main text). Thus,  $F_{r}=\num{5.4e-8} \si{\newton}$, which is two orders of magnitude lower than experimentally measured thrust $F_p=\SI{9}{\micro\newton}$  at this angle (see Fig.~2 of the main text).
Because the assumption that the solute is ejected with velocity $v_{p}$ is in fact a gross overestimation, the thrust as a result of direct ejection of the mass can be expected to be even lower by several orders of magnitude.  Thus, we conclude that the boat cannot move forward with the observed speeds because of a rocket-like propulsion mechanism due to the ejection of the dissolved mass alone.

\section{Estimate of thrust from momentum balance} 
\label{sec:thrustMB}
We compute the thrust $F_p^{MB}$ due to density currents by considering momentum balance using the experimentally-obtained velocity fields. We have
\begin{eqnarray}
        F_p^{MB} & =   \int_{y_1}^{y_2}\,\int_{z_1}^{z_2}  \rho\, v_x^2  \,\mathrm{d} y \, \mathrm{d} z \,|_{x=x_2}
  - \int_{y_1}^{y_2}\,\int_{x_1}^{x_2}  \rho\, v_z\,v_x \, \mathrm{d} y \, \mathrm{d} x  \,|_{z=z_1}  \nonumber \\ &   + \int_{z_1}^{z_2} \int_{x_1}^{x_2}  \rho\, v_y\,v_x  \, \mathrm{d} x \, \mathrm{d} z \, |_{y=- W/2} \nonumber \\&
  - \int_{z_1}^{z_2} \int_{x_1}^{x_2}  \rho\, v_y\,v_x  \, \mathrm{d} x \, \mathrm{d} z \, |_{y=+ W/2}\,.
  \label{Eq:Fp}
\end{eqnarray}
The first two terms can be directly computed from the velocity field in a vertical plane obtained with PIV measurements. The last to terms, which make equal contribution because of symmetry, can be obtained using PIV measurements in the perpendicular plane. To avoid the large number of measurements this entails, we estimate the last terms using conservation of mass. (We validate this approach by performing limited measurements in the perpendicular plane as discussed in SI Section~S\ref{sec:piv}, and further showing that those terms make relatively small contributions in SI Section~S\ref{sec:dominF}.)

Mass conservation in the control volume $\mathcal{V}$ shown in Fig.~4D in the main text can be written as
\begin{eqnarray}
        0 & =   \int_{y_1}^{y_2}\,\int_{z_1}^{z_2}  \rho\, v_x  \,\mathrm{d} y \, \mathrm{d} z \,|_{x=x_2}
  - \int_{y_1}^{y_2}\,\int_{x_1}^{x_2}  \rho\, v_z \, \mathrm{d} y \, \mathrm{d} x  \,|_{z=z_1}  \nonumber \\ &   + \int_{z_1}^{z_2} \int_{x_1}^{x_2}  \rho\, v_y  \, \mathrm{d} x \, \mathrm{d} z \, |_{y=- W/2} \nonumber \\&
  - \int_{z_1}^{z_2} \int_{x_1}^{x_2}  \rho\, v_y  \, \mathrm{d} x \, \mathrm{d} z \, |_{y=+ W/2}\,,
  \label{Eq:masscsvtion}
\end{eqnarray}
The last two terms are in fact equal in magnitude by symmetry, and we thus have, 
\begin{equation}
\int_{z_1}^{z_2} \int_{x_1}^{x_2}  \rho\, v_y  \, \mathrm{d} x \, \mathrm{d} z \, |_{y=- W/2} = - \frac{1}{2} \left( \int_{y_1}^{y_2}\,\int_{z_1}^{z_2}  \rho\, v_x  \,\mathrm{d} y \, \mathrm{d} z \,|_{x=x_2}
  - \int_{y_1}^{y_2}\,\int_{x_1}^{x_2}  \rho\, v_z \, \mathrm{d} y \, \mathrm{d} x  \,|_{z=z_1}  \right).
\label{eq:vy}
\end{equation}
The terms on the right hand side can be evaluated using the data obtained  using PIV in the $xz$-plane. 

Now, assuming that ${v_x} \simeq \bar{v_x}^{s}$, the average horizontal velocity through the sides of the control volume $\mathcal{V}$, we have $$\int_{z_1}^{z_2} \int_{x_1}^{x_2}  \rho\, v_y\,v_x  \, \mathrm{d} x \, \mathrm{d} z \, |_{y=- W/2} \simeq \bar{v_x}^{s}\int_{z_1}^{z_2} \int_{x_1}^{x_2}  \rho\, v_y  \, \mathrm{d} x \, \mathrm{d} z \, |_{y=- W/2}.$$ %In addition, we hypothesize that ${v_x} \simeq \bar{v_x}^{s}$ \green{??} the average horizontal velocity coming from the front. 
Because $\bar{v_x}^{s}$ and Eq.~[\ref{eq:vy}] can be evaluated using the velocity field obtained in the $xz$-plane, we can plug this estimate into Eq.~[\ref{Eq:Fp}], and thus obtain $F_p^{MB}$.

\section{Relative thrust contributions}
\label{sec:dominF}

\begin{figure}[h!]
\centering
\includegraphics[width=0.45\columnwidth]{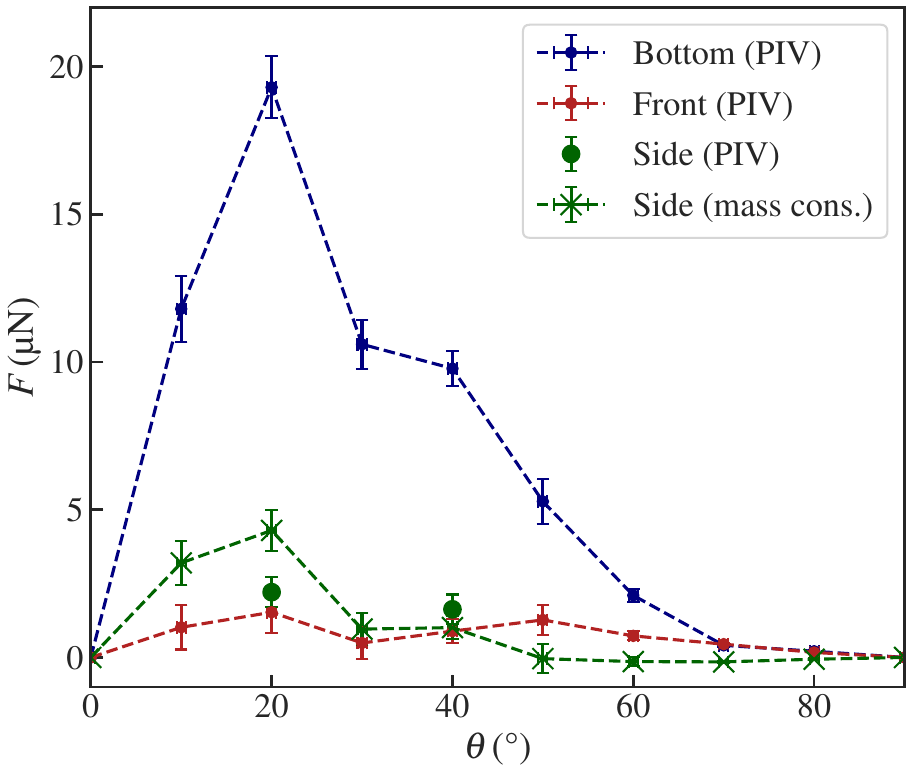}
\caption{\label{fig:piv_fpmom} Contribution of the flow coming from the front and the sides, and leaving from the bottom the control volume $\mathcal{V}$ shown in Fig.~4A in the main text, to the calculation of $F_p^{MB}$ using Eq.~[\ref{Eq:masscsvtion}]. The dominant term corresponds clearly to the momentum leaving the control volume through the bottom boundary.} 
\end{figure}

We numerically estimate the first two terms in Eq.~[\ref{Eq:Fp}] separately using the PIV data obtained in the $xz$-plane, and the third/fourth term using mass conservation as described in Section~S\ref{sec:thrustMB} to understand their relative contributions as a function of inclination angle $\theta$. Figure~\ref{fig:piv_fpmom} shows the magnitude of the force corresponding to each of these terms. It can be noted that the term corresponding to the bottom side of $\mathcal{V}$ clearly dominates, except at large $\theta$, where it is of the same order of magnitude as the other terms. This relative strength explains why the force is positive (directed opposite to the flow), and supports the simplified modeling of Section 3 in the main text, where we only consider the contribution of flux through the bottom face to evaluate a functional dependence of thrust analytically. 

In order to test the mass balance approach to evaluating the contribution of the flux entering through the side face, we performed PIV measurements in the $yz$-plane as discussed in SI Section~S\ref{sec:piv} for $\theta = 20^\circ$ and $\theta = 40^\circ$. This data is plotted in Fig.~\ref{fig:piv_fpmom}, and can be observed to be in agreement with the estimate using mass conservation. 

\newpage
\section{Inverse Friction Coefficient}

\begin{figure}[h!]
\centering
\includegraphics[width=0.48\columnwidth]{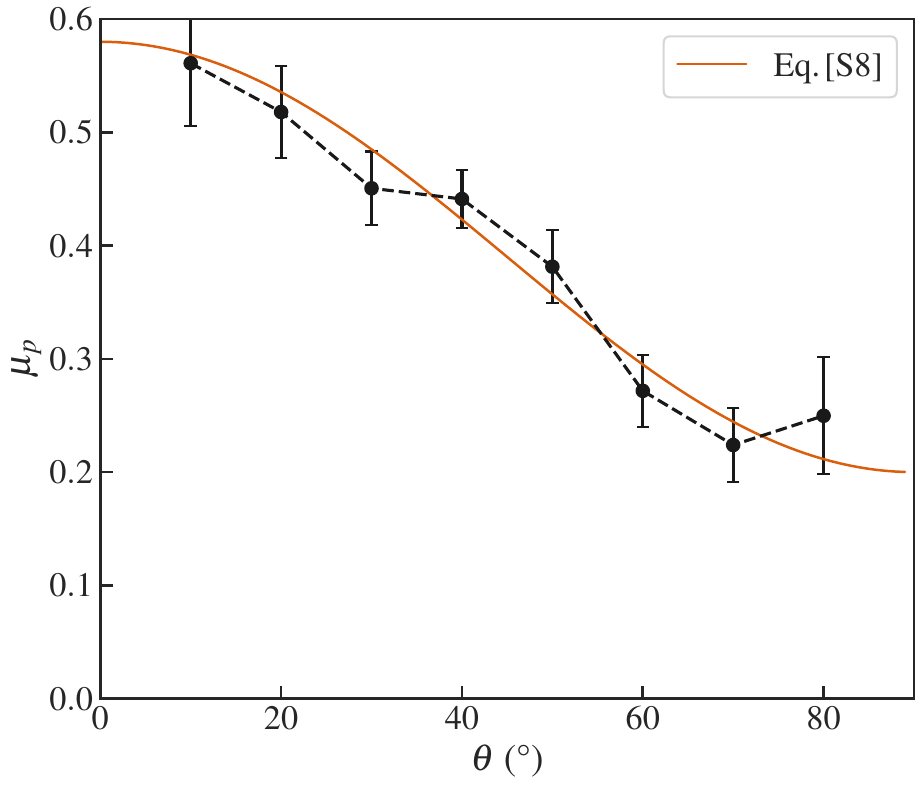}
\caption{\label{fig:vpapp} The inverse friction coefficient $\mu_p$ as a function of inclination angle $\theta$ (Black dots), and compared with fit (solid orange line) given by Eq.~[\ref{eq:mu}]. %$\mu_p= 0.2 + 0.38\,\cos^2\theta$ 
}
\end{figure}

The inverse friction coefficient $\mu_p$ is defined as $\mu_p = v_p/v_p^o$, where $v_p^{o}=\left( \frac{\rho_s \,g \, L \, \sin\theta \,\Dot{h}}{  \,\rho_f}  \right)^{1/3}$ is a reference velocity. Now, $\mu_p$ can be expected to depend on $\theta$ because the geometry of the boundary flow evolves with the inclination of the surface. 
Fig.~\ref{fig:vpapp} shows $\mu_p$ as function of $\theta$, where it can be observed to decrease with increasing inclination angle. We find that the empirical function 
\begin{equation}
\mu_p=0.2+0.38\,\cos^2\theta,
\label{eq:mu}
\end{equation}
captures the overall inclination-dependence. $\mu_p$ is observed to remain roughly the same order of magnitude while $\theta$ varies between $0^\circ$ and $90^\circ$. 

\section{Calculation of Boat Speed}
We obtain the functional form of the boat speed by combining Eq.[7] and Eq.[8] in the main text, with Eq.[\ref{eq:rrates}]. Neglecting the inclination dependency, we obtain

\begin{equation}
    U_b^{th} \approx \left( \dfrac{\delta_x^{1/2}\,L^{1/3}}{L_A^{1/2} \, \delta_c^{1/3}\,} \right)  \, \left(\dfrac{c_{sat}\rho_{sat}}{\rho_f\,(\rho_{sat}-c_{sat})}\right)^{1/3} \,  (D \,g)^{1/3} \, ,
    \label{eq:Ubth2}
\end{equation} 
where $\delta_c$ is the concentrated boundary layer thickness, and $\delta_x$ is the distance over which the gravity-driven flow is concentrated along the slab surface (see the schematic shown in Fig.~4D, in the main text). Note that the first and second fractional terms are dimensionless quantities built with lengths and density or mass concentrations, respectively. In contrast, the third factor $(D \,g)^{1/3} $ constitutes the intrinsic velocity scale in the problem. For sucrose dissolution in water $(D \,g)^{1/3}$ is of order of $\SI{1.6}{\milli \meter  \per\second}$, and for salt dissolution about $\SI{2.5}{\milli \meter \per\second}$. These speeds are surprisingly similar in order of magnitude to the corresponding measured boat speeds.

In order to obtain the explicit dependence of boat speed on its length scale, we first note that \begin{equation}
\delta_c=\left(\dfrac{Ra_c\,\nu_i\,\rho_f \,D}{g \cos \theta \, (\rho_{sat}-\rho_f)}  \right)^{1/3}\,,
\end{equation}
based on the criterion for constant Rayleigh number for solutal convection~\cite{Sullivan96,Wykes2018,Cohen2020,Sharma2021}. 
Then, if we assume that $\delta_x$ is well described by the empirical form given by  $\delta_x = \frac{1}{2}\,\cos^2 \theta\,L$ (Eq.~[3] in the main text), we obtain a simplified functional for the boat speed 
\begin{equation}
    U_b^{th} \approx   \, \left(\dfrac{D\,g\,c_{sat}\,\rho_{sat}}{\rho_f\,(\rho_{sat}-c_{sat})\, \delta_c}\right)^{1/3} L_A^{-1/2}\,L^{5/6}.
    \label{eq:Ubth3b}
\end{equation} 

The terms in the bracket depend on the properties of the dissolving solid and the bath, which are independent of the dimensions of the boat, its length $L$ and the projected boat length $L_A$. The variation of $U_b^{th}$ was observed to collapse with $L_A^{-1/2}\,L^{5/6}$ by using boats with various sizes in Fig.~5B. The projected boat length $L_A$ results mainly from the vertical projection of the boat length $L$ (see SI Section S\ref{section:LA}), thus $L_A$ increases with $L$. In our experiments for the boats of few centimeters length, the dimensions of the buoy also contribute to the value of $L_A$. However, for larger boats or for solids with uniform density as in floating ice, we can assume that correction is small, and thus $L_A \approx L$. Then, by simplifying Eq.~[\ref{eq:Ubth3b}], we find that $U_b$ increases with the dissolving slab length as
\begin{equation}
    U_b^{th} \approx   \, \left(\dfrac{D\,g\,c_{sat}\,\rho_{sat}}{\rho_f\,(\rho_{sat}-c_{sat})\, \delta_c}\right)^{1/3} \,L^{1/3}.
    \label{eq:Ubth4}
\end{equation} 
Hence, the speed can be expected to increase with the size of the dissolving body, albeit rather slowly.

%%% Add this line AFTER all your figures and tables

\section{Description of movies}

\textbf{Movie S1}: Top view of the motion of a candy boat moving in a large tank filled with fresh water viewed from the top. The boat is observed rectilinearly with speeds approaching $5$\,mm/s.  The movie frame rate has been increased x10 ($L = 7.5$\,cm; $\theta \approx 22^\circ$). The tracked position of the boat is overlaid to indicate the boat trajectory over time.\\

\textbf{Movie S2}: Top view of the motion of a candy boat moving in a large tank filled with fresh water viewed from the top. The boat is observed to move slowly and aimlessly. The movie frame rate has been increased x10 ($L = 7.5$\,cm; $\theta \approx 2^\circ$). The tracked position of the boat is overlaid to indicate the boat trajectory over time.\\

\textbf{Movie S3}: A dissolving candy boat moving in fresh water tank viewed from the side with shadowgraph imaging. The motion corresponds to real time and the movie is taken at 10 frames per second soon after the boat is launched from rest. The tank is illuminated from behind with a light source placed at the focal plane of a Fresnel lens, which produces a parallel beam and enables visualization of refraction index gradients in the solute-rich plumes. The length of the boat $L = \SI{7.5}{\centi\meter}$, and its inclination angle  $\theta=16^\circ$. \\

\textbf{Movie S4}: Side view of a fully submerged candy boat moving in a stratified fluid. The movie is speeded up by a factor 4. %The boat is composed of a $\SI{3.75}{\centi\meter}$ long candy slab glued to a hollow plastic box and ballasted with metal pieces. 
The length of the boat $L = \SI{3.75}{\centi\meter}$, and it floats at the interface between a layer of fresh water and a layer of denser salt water in a $\SI{18}{\centi\meter}$ long aquarium. Starting from the left side of the tank, the submerged boat moves rectilinearly to the right with a speed of approximately $\SI{0.1}{\centi\meter\per\second}$ over the {$\SI{14}{\second}$} time duration of the movie. \\

\textbf{Movie S5}: Top view of two boats moving in the same direction. The following boat catches up with the leading boat. ($L = 7.5$\,cm; $\theta = 33^o \pm 2^o$). The movie playback is faster by a factor of 2. \\

\textbf{Movie S7}: {Top view of two boats moving in opposite direction, which approach each other at a distance less than their widths. The boats spin around each other after coming in contact. ($L = 7.5$\,cm; $\theta = 33^o \pm 2^o$). The movie playback is faster by a factor of 2.}\\

\textbf{Movie S8}: {Top view of two boats moving in opposite direction, which approach each other at a distance slightly greater than their widths. The boats spin rapidly around each other after coming in contact side by side. ($L = 7.5$\,cm; $\theta = 33^o \pm 2^o$). The movie playback is faster by a factor of 2.}\\


\begin{thebibliography}{59}%
\makeatletter
\providecommand \@ifxundefined [1]{%
 \@ifx{#1\undefined}
}%
\providecommand \@ifnum [1]{%
 \ifnum #1\expandafter \@firstoftwo
 \else \expandafter \@secondoftwo
 \fi
}%
\providecommand \@ifx [1]{%
 \ifx #1\expandafter \@firstoftwo
 \else \expandafter \@secondoftwo
 \fi
}%
\providecommand \natexlab [1]{#1}%
\providecommand \enquote  [1]{``#1''}%
\providecommand \bibnamefont  [1]{#1}%
\providecommand \bibfnamefont [1]{#1}%
\providecommand \citenamefont [1]{#1}%
\providecommand \href@noop [0]{\@secondoftwo}%
\providecommand \href [0]{\begingroup \@sanitize@url \@href}%
\providecommand \@href[1]{\@@startlink{#1}\@@href}%
\providecommand \@@href[1]{\endgroup#1\@@endlink}%
\providecommand \@sanitize@url [0]{\catcode `\\12\catcode `\$12\catcode
  `\&12\catcode `\#12\catcode `\^12\catcode `\_12\catcode `\%12\relax}%
\providecommand \@@startlink[1]{}%
\providecommand \@@endlink[0]{}%
\providecommand \url  [0]{\begingroup\@sanitize@url \@url }%
\providecommand \@url [1]{\endgroup\@href {#1}{\urlprefix }}%
\providecommand \urlprefix  [0]{URL }%
\providecommand \Eprint [0]{\href }%
\providecommand \doibase [0]{https://doi.org/}%
\providecommand \selectlanguage [0]{\@gobble}%
\providecommand \bibinfo  [0]{\@secondoftwo}%
\providecommand \bibfield  [0]{\@secondoftwo}%
\providecommand \translation [1]{[#1]}%
\providecommand \BibitemOpen [0]{}%
\providecommand \bibitemStop [0]{}%
\providecommand \bibitemNoStop [0]{.\EOS\space}%
\providecommand \EOS [0]{\spacefactor3000\relax}%
\providecommand \BibitemShut  [1]{\csname bibitem#1\endcsname}%
\let\auto@bib@innerbib\@empty
%</preamble>
\bibitem [{\citenamefont {Marchetti}\ \emph {et~al.}(2013)\citenamefont
  {Marchetti}, \citenamefont {Joanny}, \citenamefont {Ramaswamy}, \citenamefont
  {Liverpool}, \citenamefont {Prost}, \citenamefont {Rao},\ and\ \citenamefont
  {Simha}}]{Marchetti2013}%
  \BibitemOpen
  \bibfield  {author} {\bibinfo {author} {\bibfnamefont {M.~C.}\ \bibnamefont
  {Marchetti}}, \bibinfo {author} {\bibfnamefont {J.~F.}\ \bibnamefont
  {Joanny}}, \bibinfo {author} {\bibfnamefont {S.}~\bibnamefont {Ramaswamy}},
  \bibinfo {author} {\bibfnamefont {T.~B.}\ \bibnamefont {Liverpool}}, \bibinfo
  {author} {\bibfnamefont {J.}~\bibnamefont {Prost}}, \bibinfo {author}
  {\bibfnamefont {M.}~\bibnamefont {Rao}},\ and\ \bibinfo {author}
  {\bibfnamefont {R.~A.}\ \bibnamefont {Simha}},\ }\bibfield  {title} {\bibinfo
  {title} {Hydrodynamics of soft active matter},\ }\href
  {https://doi.org/10.1103/RevModPhys.85.1143} {\bibfield  {journal} {\bibinfo
  {journal} {Rev. Mod. Phys.}\ }\textbf {\bibinfo {volume} {85}},\ \bibinfo
  {pages} {1143} (\bibinfo {year} {2013})}\BibitemShut {NoStop}%
\bibitem [{\citenamefont {Brennen}\ and\ \citenamefont
  {H.~Winet}(1977)}]{Brennen1977}%
  \BibitemOpen
  \bibfield  {author} {\bibinfo {author} {\bibfnamefont {C.}~\bibnamefont
  {Brennen}}\ and\ \bibinfo {author} {\bibfnamefont {H.}~\bibnamefont
  {H.~Winet}},\ }\bibfield  {title} {\bibinfo {title} {Fluid mechanics of
  propulsion by cilia and flagella},\ }\href
  {https://doi.org/10.1146/annurev.fl.09.010177.002011} {\bibfield  {journal}
  {\bibinfo  {journal} {Annual Review of Fluid Mechanics}\ }\textbf {\bibinfo
  {volume} {9}},\ \bibinfo {pages} {339} (\bibinfo {year} {1977})}\BibitemShut
  {NoStop}%
\bibitem [{\citenamefont {Lauga}\ and\ \citenamefont
  {Powers}(2009)}]{Lauga2009}%
  \BibitemOpen
  \bibfield  {author} {\bibinfo {author} {\bibfnamefont {E.}~\bibnamefont
  {Lauga}}\ and\ \bibinfo {author} {\bibfnamefont {T.~R.}\ \bibnamefont
  {Powers}},\ }\bibfield  {title} {\bibinfo {title} {The hydrodynamics of
  swimming microorganisms.},\ }\href
  {https://doi.org/10.1088/0034-4885/72/9/096601} {\bibfield  {journal}
  {\bibinfo  {journal} {Reports on Progress in Physics}\ }\textbf {\bibinfo
  {volume} {72}},\ \bibinfo {pages} {096601} (\bibinfo {year}
  {2009})}\BibitemShut {NoStop}%
\bibitem [{\citenamefont {Elgetti}\ \emph {et~al.}(2015)\citenamefont
  {Elgetti}, \citenamefont {Winkler},\ and\ \citenamefont
  {Gompper}}]{Elgeti2015}%
  \BibitemOpen
  \bibfield  {author} {\bibinfo {author} {\bibfnamefont {J.}~\bibnamefont
  {Elgetti}}, \bibinfo {author} {\bibfnamefont {R.~G.}\ \bibnamefont
  {Winkler}},\ and\ \bibinfo {author} {\bibfnamefont {G.}~\bibnamefont
  {Gompper}},\ }\bibfield  {title} {\bibinfo {title} {Physics of
  microswimmers--single particle motion and collective behavior: a review},\
  }\href {https://doi.org/10.1088/0034-4885/78/5/056601} {\bibfield  {journal}
  {\bibinfo  {journal} {Reports on Progress in Physics}\ }\textbf {\bibinfo
  {volume} {78}},\ \bibinfo {pages} {056601} (\bibinfo {year}
  {2015})}\BibitemShut {NoStop}%
\bibitem [{\citenamefont {Jee}\ \emph {et~al.}(2018)\citenamefont {Jee},
  \citenamefont {Cho}, \citenamefont {Granick},\ and\ \citenamefont
  {Tlusty}}]{Jee2018}%
  \BibitemOpen
  \bibfield  {author} {\bibinfo {author} {\bibfnamefont {A.-Y.}\ \bibnamefont
  {Jee}}, \bibinfo {author} {\bibfnamefont {Y.-K.}\ \bibnamefont {Cho}},
  \bibinfo {author} {\bibfnamefont {S.}~\bibnamefont {Granick}},\ and\ \bibinfo
  {author} {\bibfnamefont {T.}~\bibnamefont {Tlusty}},\ }\bibfield  {title}
  {\bibinfo {title} {Catalytic enzymes are active matter},\ }\href
  {https://doi.org/10.1073/pnas.1814180115} {\bibfield  {journal} {\bibinfo
  {journal} {Proceedings of the National Academy of Sciences}\ }\textbf
  {\bibinfo {volume} {115}},\ \bibinfo {pages} {E10812} (\bibinfo {year}
  {2018})},\ \Eprint
  {https://arxiv.org/abs/https://www.pnas.org/content/115/46/E10812.full.pdf}
  {https://www.pnas.org/content/115/46/E10812.full.pdf} \BibitemShut {NoStop}%
\bibitem [{\citenamefont {Dorbolo}\ \emph {et~al.}(2005)\citenamefont
  {Dorbolo}, \citenamefont {Volfson}, \citenamefont {Tsimring},\ and\
  \citenamefont {Kudrolli}}]{dorbolo2005}%
  \BibitemOpen
  \bibfield  {author} {\bibinfo {author} {\bibfnamefont {S.}~\bibnamefont
  {Dorbolo}}, \bibinfo {author} {\bibfnamefont {D.}~\bibnamefont {Volfson}},
  \bibinfo {author} {\bibfnamefont {L.}~\bibnamefont {Tsimring}},\ and\
  \bibinfo {author} {\bibfnamefont {A.}~\bibnamefont {Kudrolli}},\ }\bibfield
  {title} {\bibinfo {title} {Dynamics of a bouncing dimer},\ }\href
  {https://doi.org/10.1103/PhysRevLett.95.044101} {\bibfield  {journal}
  {\bibinfo  {journal} {Phys. Rev. Lett.}\ }\textbf {\bibinfo {volume} {95}},\
  \bibinfo {pages} {044101} (\bibinfo {year} {2005})}\BibitemShut {NoStop}%
\bibitem [{\citenamefont {Kudrolli}\ \emph {et~al.}(2008)\citenamefont
  {Kudrolli}, \citenamefont {Lumay}, \citenamefont {Volfson},\ and\
  \citenamefont {Tsimring}}]{kudrolli2008}%
  \BibitemOpen
  \bibfield  {author} {\bibinfo {author} {\bibfnamefont {A.}~\bibnamefont
  {Kudrolli}}, \bibinfo {author} {\bibfnamefont {G.}~\bibnamefont {Lumay}},
  \bibinfo {author} {\bibfnamefont {D.}~\bibnamefont {Volfson}},\ and\ \bibinfo
  {author} {\bibfnamefont {L.~S.}\ \bibnamefont {Tsimring}},\ }\bibfield
  {title} {\bibinfo {title} {Swarming and swirling in self-propelled polar
  granular rods},\ }\href {https://doi.org/10.1103/PhysRevLett.100.058001}
  {\bibfield  {journal} {\bibinfo  {journal} {Phys. Rev. Lett.}\ }\textbf
  {\bibinfo {volume} {100}},\ \bibinfo {pages} {058001} (\bibinfo {year}
  {2008})}\BibitemShut {NoStop}%
\bibitem [{\citenamefont {Golestanian}\ \emph {et~al.}(2005)\citenamefont
  {Golestanian}, \citenamefont {Liverpool},\ and\ \citenamefont
  {Ajdari}}]{Golestanian2005}%
  \BibitemOpen
  \bibfield  {author} {\bibinfo {author} {\bibfnamefont {R.}~\bibnamefont
  {Golestanian}}, \bibinfo {author} {\bibfnamefont {T.~B.}\ \bibnamefont
  {Liverpool}},\ and\ \bibinfo {author} {\bibfnamefont {A.}~\bibnamefont
  {Ajdari}},\ }\bibfield  {title} {\bibinfo {title} {Propulsion of a molecular
  machine by asymmetric distribution of reaction products},\ }\href
  {https://doi.org/10.1103/PhysRevLett.94.220801} {\bibfield  {journal}
  {\bibinfo  {journal} {Phys. Rev. Lett.}\ }\textbf {\bibinfo {volume} {94}},\
  \bibinfo {pages} {220801} (\bibinfo {year} {2005})}\BibitemShut {NoStop}%
\bibitem [{\citenamefont {Izri}\ \emph {et~al.}(2014)\citenamefont {Izri},
  \citenamefont {van~der Linden}, \citenamefont {Michelin},\ and\ \citenamefont
  {Dauchot}}]{Izri2014}%
  \BibitemOpen
  \bibfield  {author} {\bibinfo {author} {\bibfnamefont {Z.}~\bibnamefont
  {Izri}}, \bibinfo {author} {\bibfnamefont {M.~N.}\ \bibnamefont {van~der
  Linden}}, \bibinfo {author} {\bibfnamefont {S.}~\bibnamefont {Michelin}},\
  and\ \bibinfo {author} {\bibfnamefont {O.}~\bibnamefont {Dauchot}},\
  }\bibfield  {title} {\bibinfo {title} {Self-propulsion of pure water droplets
  by spontaneous marangoni-stress-driven motion},\ }\href
  {https://doi.org/10.1103/PhysRevLett.113.248302} {\bibfield  {journal}
  {\bibinfo  {journal} {Phys. Rev. Lett.}\ }\textbf {\bibinfo {volume} {113}},\
  \bibinfo {pages} {248302} (\bibinfo {year} {2014})}\BibitemShut {NoStop}%
\bibitem [{\citenamefont {Maass}\ \emph {et~al.}(2016)\citenamefont {Maass},
  \citenamefont {Kr\"uger}, \citenamefont {Herminghaus},\ and\ \citenamefont
  {Bahr}}]{Maas2016}%
  \BibitemOpen
  \bibfield  {author} {\bibinfo {author} {\bibfnamefont {C.~C.}\ \bibnamefont
  {Maass}}, \bibinfo {author} {\bibfnamefont {C.}~\bibnamefont {Kr\"uger}},
  \bibinfo {author} {\bibfnamefont {S.}~\bibnamefont {Herminghaus}},\ and\
  \bibinfo {author} {\bibfnamefont {C.}~\bibnamefont {Bahr}},\ }\bibfield
  {title} {\bibinfo {title} {Swimming droplets},\ }\href
  {https://doi.org/10.1146/annurev-conmatphys-031115-011517} {\bibfield
  {journal} {\bibinfo  {journal} {Annu. Rev. Condens. Matter Phys.}\ }\textbf
  {\bibinfo {volume} {7}},\ \bibinfo {pages} {171} (\bibinfo {year}
  {2016})}\BibitemShut {NoStop}%
\bibitem [{\citenamefont {S.Ryazantsev}\ \emph {et~al.}(2017)\citenamefont
  {S.Ryazantsev}, \citenamefont {G.Velarde}, \citenamefont {Rubio},
  \citenamefont {Guzm\'an}, \citenamefont {Ortega},\ and\ \citenamefont
  {L\'opez}}]{Ryazantsev2017}%
  \BibitemOpen
  \bibfield  {author} {\bibinfo {author} {\bibfnamefont {Y.}~\bibnamefont
  {S.Ryazantsev}}, \bibinfo {author} {\bibfnamefont {M.}~\bibnamefont
  {G.Velarde}}, \bibinfo {author} {\bibfnamefont {R.~G.}\ \bibnamefont
  {Rubio}}, \bibinfo {author} {\bibfnamefont {E.}~\bibnamefont {Guzm\'an}},
  \bibinfo {author} {\bibfnamefont {F.}~\bibnamefont {Ortega}},\ and\ \bibinfo
  {author} {\bibfnamefont {P.}~\bibnamefont {L\'opez}},\ }\bibfield  {title}
  {\bibinfo {title} {Thermo- and soluto-capillarity: Passive and active
  drops},\ }\href {https://doi.org/https://doi.org/10.1016/j.cis.2017.07.025}
  {\bibfield  {journal} {\bibinfo  {journal} {Advances in Colloid and Interface
  Science}\ }\textbf {\bibinfo {volume} {247}},\ \bibinfo {pages} {52}
  (\bibinfo {year} {2017})}\BibitemShut {NoStop}%
\bibitem [{\citenamefont {Reichert}\ \emph {et~al.}(2021)\citenamefont
  {Reichert}, \citenamefont {Cam}, \citenamefont {Saint-Jalmes},\ and\
  \citenamefont {Pucci}}]{Reichert2021}%
  \BibitemOpen
  \bibfield  {author} {\bibinfo {author} {\bibfnamefont {B.}~\bibnamefont
  {Reichert}}, \bibinfo {author} {\bibfnamefont {J.-B.~L.}\ \bibnamefont
  {Cam}}, \bibinfo {author} {\bibfnamefont {A.}~\bibnamefont {Saint-Jalmes}},\
  and\ \bibinfo {author} {\bibfnamefont {G.}~\bibnamefont {Pucci}},\ }\bibfield
   {title} {\bibinfo {title} {Self-propulsion of a volatile drop on the surface
  of an immiscible liquid bath},\ }\href
  {https://doi.org/10.1103/PhysRevLett.127.144501} {\bibfield  {journal}
  {\bibinfo  {journal} {Phys. Rev. Lett.}\ }\textbf {\bibinfo {volume} {127}},\
  \bibinfo {pages} {144501} (\bibinfo {year} {2021})}\BibitemShut {NoStop}%
\bibitem [{\citenamefont {Linke}\ \emph {et~al.}(2006)\citenamefont {Linke},
  \citenamefont {Alem\'an}, \citenamefont {Melling}, \citenamefont {Taormina},
  \citenamefont {Francis}, \citenamefont {Dow-Hygelund}, \citenamefont
  {Narayanan}, \citenamefont {Taylor},\ and\ \citenamefont
  {Stout}}]{Linke2006}%
  \BibitemOpen
  \bibfield  {author} {\bibinfo {author} {\bibfnamefont {H.}~\bibnamefont
  {Linke}}, \bibinfo {author} {\bibfnamefont {B.~J.}\ \bibnamefont {Alem\'an}},
  \bibinfo {author} {\bibfnamefont {L.~D.}\ \bibnamefont {Melling}}, \bibinfo
  {author} {\bibfnamefont {M.~J.}\ \bibnamefont {Taormina}}, \bibinfo {author}
  {\bibfnamefont {M.~J.}\ \bibnamefont {Francis}}, \bibinfo {author}
  {\bibfnamefont {C.~C.}\ \bibnamefont {Dow-Hygelund}}, \bibinfo {author}
  {\bibfnamefont {V.}~\bibnamefont {Narayanan}}, \bibinfo {author}
  {\bibfnamefont {R.~P.}\ \bibnamefont {Taylor}},\ and\ \bibinfo {author}
  {\bibfnamefont {A.}~\bibnamefont {Stout}},\ }\bibfield  {title} {\bibinfo
  {title} {Self-propelled leidenfrost droplets},\ }\href
  {https://doi.org/10.1103/PhysRevLett.96.154502} {\bibfield  {journal}
  {\bibinfo  {journal} {Phys. Rev. Lett.}\ }\textbf {\bibinfo {volume} {96}},\
  \bibinfo {pages} {154502} (\bibinfo {year} {2006})}\BibitemShut {NoStop}%
\bibitem [{\citenamefont {Lagubeau}\ \emph {et~al.}(2011)\citenamefont
  {Lagubeau}, \citenamefont {Merrer}, \citenamefont {Clanet},\ and\
  \citenamefont {Qu\'er\'e}}]{Lagubeau2011}%
  \BibitemOpen
  \bibfield  {author} {\bibinfo {author} {\bibfnamefont {G.}~\bibnamefont
  {Lagubeau}}, \bibinfo {author} {\bibfnamefont {M.~L.}\ \bibnamefont
  {Merrer}}, \bibinfo {author} {\bibfnamefont {C.}~\bibnamefont {Clanet}},\
  and\ \bibinfo {author} {\bibfnamefont {D.}~\bibnamefont {Qu\'er\'e}},\
  }\bibfield  {title} {\bibinfo {title} {A liquid droplet placed on a hot
  surface can levitate, and moreover, self-propel if the surface is textured.
  solids can similarly self-propel},\ }\href
  {https://doi.org/10.1038/nphys1925} {\bibfield  {journal} {\bibinfo
  {journal} {Nature Physics}\ }\textbf {\bibinfo {volume} {7}},\ \bibinfo
  {pages} {395} (\bibinfo {year} {2011})}\BibitemShut {NoStop}%
\bibitem [{\citenamefont {Gauthier}\ \emph {et~al.}(2019)\citenamefont
  {Gauthier}, \citenamefont {Diddens}, \citenamefont {Proville}, \citenamefont
  {Lohse},\ and\ \citenamefont {van~der Meer}}]{Gauthier2019}%
  \BibitemOpen
  \bibfield  {author} {\bibinfo {author} {\bibfnamefont {A.}~\bibnamefont
  {Gauthier}}, \bibinfo {author} {\bibfnamefont {C.}~\bibnamefont {Diddens}},
  \bibinfo {author} {\bibfnamefont {R.}~\bibnamefont {Proville}}, \bibinfo
  {author} {\bibfnamefont {D.}~\bibnamefont {Lohse}},\ and\ \bibinfo {author}
  {\bibfnamefont {D.}~\bibnamefont {van~der Meer}},\ }\bibfield  {title}
  {\bibinfo {title} {Self-propulsion of inverse leidenfrost drops on a
  cryogenic bath},\ }\href {https://doi.org/10.1073/pnas.1812288116} {\bibfield
   {journal} {\bibinfo  {journal} {Proceedings of the National Academy of
  Sciences}\ }\textbf {\bibinfo {volume} {116}},\ \bibinfo {pages} {1174}
  (\bibinfo {year} {2019})}\BibitemShut {NoStop}%
\bibitem [{\citenamefont {{Allshouse}}\ \emph {et~al.}(2010)\citenamefont
  {{Allshouse}}, \citenamefont {{Barad}},\ and\ \citenamefont
  {{Peacock}}}]{Allshouse2010}%
  \BibitemOpen
  \bibfield  {author} {\bibinfo {author} {\bibfnamefont {M.~R.}\ \bibnamefont
  {{Allshouse}}}, \bibinfo {author} {\bibfnamefont {M.~F.}\ \bibnamefont
  {{Barad}}},\ and\ \bibinfo {author} {\bibfnamefont {T.}~\bibnamefont
  {{Peacock}}},\ }\bibfield  {title} {\bibinfo {title} {{Propulsion generated
  by diffusion-driven flow}},\ }\href {https://doi.org/10.1038/nphys1686}
  {\bibfield  {journal} {\bibinfo  {journal} {Nature Physics}\ }\textbf
  {\bibinfo {volume} {6}},\ \bibinfo {pages} {516} (\bibinfo {year}
  {2010})}\BibitemShut {NoStop}%
\bibitem [{\citenamefont {Mercier}\ \emph {et~al.}(2014)\citenamefont
  {Mercier}, \citenamefont {Ardekani}, \citenamefont {Allshouse}, \citenamefont
  {Doyle},\ and\ \citenamefont {Peacock}}]{Mercier2014}%
  \BibitemOpen
  \bibfield  {author} {\bibinfo {author} {\bibfnamefont {M.~J.}\ \bibnamefont
  {Mercier}}, \bibinfo {author} {\bibfnamefont {A.~M.}\ \bibnamefont
  {Ardekani}}, \bibinfo {author} {\bibfnamefont {M.~R.}\ \bibnamefont
  {Allshouse}}, \bibinfo {author} {\bibfnamefont {B.}~\bibnamefont {Doyle}},\
  and\ \bibinfo {author} {\bibfnamefont {T.}~\bibnamefont {Peacock}},\
  }\bibfield  {title} {\bibinfo {title} {Self-propulsion of immersed objects
  via natural convection},\ }\href
  {https://doi.org/10.1103/PhysRevLett.112.204501} {\bibfield  {journal}
  {\bibinfo  {journal} {Phys. Rev. Lett.}\ }\textbf {\bibinfo {volume} {112}},\
  \bibinfo {pages} {204501} (\bibinfo {year} {2014})}\BibitemShut {NoStop}%
\bibitem [{\citenamefont {Sullivan}\ \emph {et~al.}(1996)\citenamefont
  {Sullivan}, \citenamefont {Liu},\ and\ \citenamefont {Ecke}}]{Sullivan96}%
  \BibitemOpen
  \bibfield  {author} {\bibinfo {author} {\bibfnamefont {T.~S.}\ \bibnamefont
  {Sullivan}}, \bibinfo {author} {\bibfnamefont {Y.}~\bibnamefont {Liu}},\ and\
  \bibinfo {author} {\bibfnamefont {R.~E.}\ \bibnamefont {Ecke}},\ }\bibfield
  {title} {\bibinfo {title} {Turbulent solutal convection and surface
  patterning in solid dissolution},\ }\href
  {https://doi.org/10.1103/PhysRevE.54.486} {\bibfield  {journal} {\bibinfo
  {journal} {Phys. Rev. E}\ }\textbf {\bibinfo {volume} {54}},\ \bibinfo
  {pages} {486} (\bibinfo {year} {1996})}\BibitemShut {NoStop}%
\bibitem [{\citenamefont {{Davies~Wykes}}\ \emph {et~al.}(2018)\citenamefont
  {{Davies~Wykes}}, \citenamefont {Huang}, \citenamefont {Hajjar},\ and\
  \citenamefont {Ristroph}}]{Wykes2018}%
  \BibitemOpen
  \bibfield  {author} {\bibinfo {author} {\bibfnamefont {M.~S.}\ \bibnamefont
  {{Davies~Wykes}}}, \bibinfo {author} {\bibfnamefont {J.~M.}\ \bibnamefont
  {Huang}}, \bibinfo {author} {\bibfnamefont {G.~A.}\ \bibnamefont {Hajjar}},\
  and\ \bibinfo {author} {\bibfnamefont {L.}~\bibnamefont {Ristroph}},\
  }\bibfield  {title} {\bibinfo {title} {{Self-sculpting of a dissolvable body
  due to gravitational convection}},\ }\href
  {https://doi.org/10.1103/PhysRevFluids.3.043801} {\bibfield  {journal}
  {\bibinfo  {journal} {Physical Review Fluids}\ }\textbf {\bibinfo {volume}
  {3}},\ \bibinfo {pages} {043801} (\bibinfo {year} {2018})}\BibitemShut
  {NoStop}%
\bibitem [{\citenamefont {Philippi}\ \emph {et~al.}(2019)\citenamefont
  {Philippi}, \citenamefont {Berhanu}, \citenamefont {Derr},\ and\
  \citenamefont {Courrech~du Pont}}]{Philippi2019}%
  \BibitemOpen
  \bibfield  {author} {\bibinfo {author} {\bibfnamefont {J.}~\bibnamefont
  {Philippi}}, \bibinfo {author} {\bibfnamefont {M.}~\bibnamefont {Berhanu}},
  \bibinfo {author} {\bibfnamefont {J.}~\bibnamefont {Derr}},\ and\ \bibinfo
  {author} {\bibfnamefont {S.}~\bibnamefont {Courrech~du Pont}},\ }\bibfield
  {title} {\bibinfo {title} {Solutal convection induced by dissolution},\
  }\href {https://doi.org/10.1103/PhysRevFluids.4.103801} {\bibfield  {journal}
  {\bibinfo  {journal} {Phys. Rev. Fluids}\ }\textbf {\bibinfo {volume} {4}},\
  \bibinfo {pages} {103801} (\bibinfo {year} {2019})}\BibitemShut {NoStop}%
\bibitem [{\citenamefont {Cohen}\ \emph {et~al.}(2020)\citenamefont {Cohen},
  \citenamefont {Berhanu}, \citenamefont {Derr},\ and\ \citenamefont
  {Courrech~du Pont}}]{Cohen2020}%
  \BibitemOpen
  \bibfield  {author} {\bibinfo {author} {\bibfnamefont {C.}~\bibnamefont
  {Cohen}}, \bibinfo {author} {\bibfnamefont {M.}~\bibnamefont {Berhanu}},
  \bibinfo {author} {\bibfnamefont {J.}~\bibnamefont {Derr}},\ and\ \bibinfo
  {author} {\bibfnamefont {S.}~\bibnamefont {Courrech~du Pont}},\ }\bibfield
  {title} {\bibinfo {title} {Buoyancy-driven dissolution of inclined blocks:
  Erosion rate and pattern formation},\ }\href
  {https://doi.org/10.1103/PhysRevFluids.5.053802} {\bibfield  {journal}
  {\bibinfo  {journal} {Phys. Rev. Fluids}\ }\textbf {\bibinfo {volume} {5}},\
  \bibinfo {pages} {053802} (\bibinfo {year} {2020})}\BibitemShut {NoStop}%
\bibitem [{\citenamefont {Huang}\ \emph {et~al.}(2020)\citenamefont {Huang},
  \citenamefont {Tong}, \citenamefont {Shelley},\ and\ \citenamefont
  {Ristroph}}]{Huang2020}%
  \BibitemOpen
  \bibfield  {author} {\bibinfo {author} {\bibfnamefont {J.~M.}\ \bibnamefont
  {Huang}}, \bibinfo {author} {\bibfnamefont {J.}~\bibnamefont {Tong}},
  \bibinfo {author} {\bibfnamefont {M.}~\bibnamefont {Shelley}},\ and\ \bibinfo
  {author} {\bibfnamefont {L.}~\bibnamefont {Ristroph}},\ }\bibfield  {title}
  {\bibinfo {title} {{Ultra-sharp pinnacles sculpted by natural convective
  dissolution}},\ }\href {https://doi.org/10.1073/pnas.2001524117} {\bibfield
  {journal} {\bibinfo  {journal} {PNAS}\ }\textbf {\bibinfo {volume} {117}},\
  \bibinfo {pages} {23339} (\bibinfo {year} {2020})}\BibitemShut {NoStop}%
\bibitem [{\citenamefont {Dorbolo}\ \emph {et~al.}(2016)\citenamefont
  {Dorbolo}, \citenamefont {Adami}, \citenamefont {Dubois}, \citenamefont
  {Caps}, \citenamefont {Vandewalle},\ and\ \citenamefont
  {Darbois-Texier}}]{Dorbolo2016}%
  \BibitemOpen
  \bibfield  {author} {\bibinfo {author} {\bibfnamefont {S.}~\bibnamefont
  {Dorbolo}}, \bibinfo {author} {\bibfnamefont {N.}~\bibnamefont {Adami}},
  \bibinfo {author} {\bibfnamefont {C.}~\bibnamefont {Dubois}}, \bibinfo
  {author} {\bibfnamefont {H.}~\bibnamefont {Caps}}, \bibinfo {author}
  {\bibfnamefont {N.}~\bibnamefont {Vandewalle}},\ and\ \bibinfo {author}
  {\bibfnamefont {B.}~\bibnamefont {Darbois-Texier}},\ }\bibfield  {title}
  {\bibinfo {title} {Rotation of melting ice disks due to melt fluid flow},\
  }\href {https://doi.org/10.1103/PhysRevE.93.033112} {\bibfield  {journal}
  {\bibinfo  {journal} {Phys. Rev. E}\ }\textbf {\bibinfo {volume} {93}},\
  \bibinfo {pages} {033112} (\bibinfo {year} {2016})}\BibitemShut {NoStop}%
\bibitem [{\citenamefont {Chamolly}\ and\ \citenamefont
  {Lauga}(2019)}]{Chamolly2019}%
  \BibitemOpen
  \bibfield  {author} {\bibinfo {author} {\bibfnamefont {A.}~\bibnamefont
  {Chamolly}}\ and\ \bibinfo {author} {\bibfnamefont {E.}~\bibnamefont
  {Lauga}},\ }\bibfield  {title} {\bibinfo {title} {Stochastic dynamics of
  dissolving active particles},\ }\bibfield  {journal} {\bibinfo  {journal}
  {The European Physical Journal E}\ }\textbf {\bibinfo {volume} {42}},\ \href
  {https://doi.org/10.1140/epje/i2019-11854-3} {10.1140/epje/i2019-11854-3}
  (\bibinfo {year} {2019})\BibitemShut {NoStop}%
\bibitem [{\citenamefont {Mountain}(1980)}]{Mountain1980}%
  \BibitemOpen
  \bibfield  {author} {\bibinfo {author} {\bibfnamefont {D.~G.}\ \bibnamefont
  {Mountain}},\ }\bibfield  {title} {\bibinfo {title} {On predicting ice
  drift},\ }\href {https://doi.org/10.1016/0165-232X(80)90055-5} {\bibfield
  {journal} {\bibinfo  {journal} {Cold Regions Science and Technology}\
  }\textbf {\bibinfo {volume} {1}},\ \bibinfo {pages} {273} (\bibinfo {year}
  {1980})}\BibitemShut {NoStop}%
\bibitem [{\citenamefont {Andersson}\ \emph {et~al.}(2016)\citenamefont
  {Andersson}, \citenamefont {Scibilia},\ and\ \citenamefont
  {Imsland}}]{Anderson2016}%
  \BibitemOpen
  \bibfield  {author} {\bibinfo {author} {\bibfnamefont {L.~E.}\ \bibnamefont
  {Andersson}}, \bibinfo {author} {\bibfnamefont {F.}~\bibnamefont
  {Scibilia}},\ and\ \bibinfo {author} {\bibfnamefont {L.}~\bibnamefont
  {Imsland}},\ }\bibfield  {title} {\bibinfo {title} {An estimation-forecast
  set-up for iceberg drift prediction},\ }\href
  {https://doi.org/https://doi.org/10.1016/j.coldregions.2016.08.001}
  {\bibfield  {journal} {\bibinfo  {journal} {Cold Regions Science and
  Technology}\ }\textbf {\bibinfo {volume} {131}},\ \bibinfo {pages} {88}
  (\bibinfo {year} {2016})}\BibitemShut {NoStop}%
\bibitem [{\citenamefont {Feltham}(2008)}]{Feltham2008}%
  \BibitemOpen
  \bibfield  {author} {\bibinfo {author} {\bibfnamefont {D.}~\bibnamefont
  {Feltham}},\ }\bibfield  {title} {\bibinfo {title} {Sea ice rheology},\
  }\href {https://doi.org/https://doi.org/10.1016/j.coldregions.2016.08.001}
  {\bibfield  {journal} {\bibinfo  {journal} {Annual Review of Fluid
  Mechanics}\ }\textbf {\bibinfo {volume} {40}},\ \bibinfo {pages} {91}
  (\bibinfo {year} {2008})}\BibitemShut {NoStop}%
\bibitem [{\citenamefont {Alberello}\ \emph {et~al.}(2020)\citenamefont
  {Alberello}, \citenamefont {Bennetts}, \citenamefont {Heil}, \citenamefont
  {Eayrs}, \citenamefont {Vichi}, \citenamefont {MacHutchon}, \citenamefont
  {Onorato},\ and\ \citenamefont {Toffoli}}]{Alberello2020}%
  \BibitemOpen
  \bibfield  {author} {\bibinfo {author} {\bibfnamefont {A.}~\bibnamefont
  {Alberello}}, \bibinfo {author} {\bibfnamefont {L.}~\bibnamefont {Bennetts}},
  \bibinfo {author} {\bibfnamefont {P.}~\bibnamefont {Heil}}, \bibinfo {author}
  {\bibfnamefont {C.}~\bibnamefont {Eayrs}}, \bibinfo {author} {\bibfnamefont
  {M.}~\bibnamefont {Vichi}}, \bibinfo {author} {\bibfnamefont
  {K.}~\bibnamefont {MacHutchon}}, \bibinfo {author} {\bibfnamefont
  {M.}~\bibnamefont {Onorato}},\ and\ \bibinfo {author} {\bibfnamefont
  {A.}~\bibnamefont {Toffoli}},\ }\bibfield  {title} {\bibinfo {title} {Drift
  of pancake ice floes in the winter antarctic marginal ice zone during polar
  cyclones},\ }\href {https://doi.org/10.1029/2019JC015418} {\bibfield
  {journal} {\bibinfo  {journal} {Journal of Geophysical Research: Oceans}\
  }\textbf {\bibinfo {volume} {125}},\ \bibinfo {pages} {e2019JC015418.}
  (\bibinfo {year} {2020})}\BibitemShut {NoStop}%
\bibitem [{\citenamefont {Nakata}\ \emph {et~al.}(1997)\citenamefont {Nakata},
  \citenamefont {Iguchi}, \citenamefont {Ose}, \citenamefont {Kuboyama},
  \citenamefont {Ishii},\ and\ \citenamefont {Yoshikawa}}]{Nakata1997}%
  \BibitemOpen
  \bibfield  {author} {\bibinfo {author} {\bibfnamefont {S.}~\bibnamefont
  {Nakata}}, \bibinfo {author} {\bibfnamefont {Y.}~\bibnamefont {Iguchi}},
  \bibinfo {author} {\bibfnamefont {S.}~\bibnamefont {Ose}}, \bibinfo {author}
  {\bibfnamefont {M.}~\bibnamefont {Kuboyama}}, \bibinfo {author}
  {\bibfnamefont {T.}~\bibnamefont {Ishii}},\ and\ \bibinfo {author}
  {\bibfnamefont {K.}~\bibnamefont {Yoshikawa}},\ }\bibfield  {title} {\bibinfo
  {title} {Self-rotation of a camphor scraping on water: new insight into the
  old problem},\ }\href@noop {} {\bibfield  {journal} {\bibinfo  {journal}
  {Langmuir}\ }\textbf {\bibinfo {volume} {13}},\ \bibinfo {pages} {4454}
  (\bibinfo {year} {1997})}\BibitemShut {NoStop}%
\bibitem [{\citenamefont {Nagayama}\ \emph {et~al.}(2004)\citenamefont
  {Nagayama}, \citenamefont {Nakata}, \citenamefont {Doi},\ and\ \citenamefont
  {Hayashima}}]{Nagayama2004}%
  \BibitemOpen
  \bibfield  {author} {\bibinfo {author} {\bibfnamefont {M.}~\bibnamefont
  {Nagayama}}, \bibinfo {author} {\bibfnamefont {S.}~\bibnamefont {Nakata}},
  \bibinfo {author} {\bibfnamefont {Y.}~\bibnamefont {Doi}},\ and\ \bibinfo
  {author} {\bibfnamefont {Y.}~\bibnamefont {Hayashima}},\ }\bibfield  {title}
  {\bibinfo {title} {A theoretical and experimental study on the unidirectional
  motion of a camphor disk},\ }\href@noop {} {\bibfield  {journal} {\bibinfo
  {journal} {Physica D: Nonlinear Phenomena}\ }\textbf {\bibinfo {volume}
  {194}},\ \bibinfo {pages} {151} (\bibinfo {year} {2004})}\BibitemShut
  {NoStop}%
\bibitem [{\citenamefont {Biswas}\ \emph {et~al.}(2020)\citenamefont {Biswas},
  \citenamefont {Cruz}, \citenamefont {Parmananda},\ and\ \citenamefont
  {Das}}]{Biswas2020}%
  \BibitemOpen
  \bibfield  {author} {\bibinfo {author} {\bibfnamefont {A.}~\bibnamefont
  {Biswas}}, \bibinfo {author} {\bibfnamefont {J.}~\bibnamefont {Cruz}},
  \bibinfo {author} {\bibfnamefont {P.}~\bibnamefont {Parmananda}},\ and\
  \bibinfo {author} {\bibfnamefont {D.}~\bibnamefont {Das}},\ }\bibfield
  {title} {\bibinfo {title} {First passage of an active particle in the
  presence of passive crowders},\ }\href {https://doi.org/10.1039/D0SM00350F}
  {\bibfield  {journal} {\bibinfo  {journal} {Soft Matter}\ }\textbf {\bibinfo
  {volume} {16}},\ \bibinfo {pages} {6138} (\bibinfo {year}
  {2020})}\BibitemShut {NoStop}%
\bibitem [{\citenamefont {Lide}(2004)}]{Handbook}%
  \BibitemOpen
  \bibinfo {editor} {\bibfnamefont {D.~R.}\ \bibnamefont {Lide}},\ ed.,\
  \href@noop {} {\emph {\bibinfo {title} {The Handbook of Chemistry and
  Physics}}}\ (\bibinfo  {publisher} {CRC Press},\ \bibinfo {year}
  {2004})\BibitemShut {NoStop}%
\bibitem [{\citenamefont {Vella}\ and\ \citenamefont
  {Mahadevan}(2005)}]{Vella2005}%
  \BibitemOpen
  \bibfield  {author} {\bibinfo {author} {\bibfnamefont {D.}~\bibnamefont
  {Vella}}\ and\ \bibinfo {author} {\bibfnamefont {L.}~\bibnamefont
  {Mahadevan}},\ }\bibfield  {title} {\bibinfo {title} {The “cheerios
  effect”},\ }\href {https://doi.org/10.1119/1.1898523} {\bibfield  {journal}
  {\bibinfo  {journal} {American Journal of Physics}\ }\textbf {\bibinfo
  {volume} {73}},\ \bibinfo {pages} {817} (\bibinfo {year} {2005})}\BibitemShut
  {NoStop}%
\bibitem [{\citenamefont {Dalbe}\ \emph {et~al.}(2011)\citenamefont {Dalbe},
  \citenamefont {Cosic}, \citenamefont {Berhanu},\ and\ \citenamefont
  {Kudrolli}}]{Dalbe2011}%
  \BibitemOpen
  \bibfield  {author} {\bibinfo {author} {\bibfnamefont {M.~J.}\ \bibnamefont
  {Dalbe}}, \bibinfo {author} {\bibfnamefont {D.}~\bibnamefont {Cosic}},
  \bibinfo {author} {\bibfnamefont {M.}~\bibnamefont {Berhanu}},\ and\ \bibinfo
  {author} {\bibfnamefont {A.}~\bibnamefont {Kudrolli}},\ }\bibfield  {title}
  {\bibinfo {title} {Aggregation of frictional particles due to capillary
  attraction},\ }\href {https://doi.org/10.1103/PhysRevE.83.051403} {\bibfield
  {journal} {\bibinfo  {journal} {Physical Review E}\ }\textbf {\bibinfo
  {volume} {83}},\ \bibinfo {pages} {051403} (\bibinfo {year}
  {2011})}\BibitemShut {NoStop}%
\bibitem [{\citenamefont {Gazzola}\ \emph {et~al.}(2014)\citenamefont
  {Gazzola}, \citenamefont {Argentina},\ and\ \citenamefont
  {Mahadevan}}]{Gazzola2014}%
  \BibitemOpen
  \bibfield  {author} {\bibinfo {author} {\bibfnamefont {M.}~\bibnamefont
  {Gazzola}}, \bibinfo {author} {\bibfnamefont {M.}~\bibnamefont {Argentina}},\
  and\ \bibinfo {author} {\bibfnamefont {L.}~\bibnamefont {Mahadevan}},\
  }\bibfield  {title} {\bibinfo {title} {Scaling macroscopic aquatic
  locomotion},\ }\href@noop {} {\bibfield  {journal} {\bibinfo  {journal}
  {Nature Physics}\ }\textbf {\bibinfo {volume} {10}},\ \bibinfo {pages} {758}
  (\bibinfo {year} {2014})}\BibitemShut {NoStop}%
\bibitem [{\citenamefont {Gazzola}\ and\ \citenamefont
  {Argentina}(2015)}]{Gazzola2015}%
  \BibitemOpen
  \bibfield  {author} {\bibinfo {author} {\bibfnamefont {M.}~\bibnamefont
  {Gazzola}}\ and\ \bibinfo {author} {\bibfnamefont {M.}~\bibnamefont
  {Argentina}},\ }\bibfield  {title} {\bibinfo {title} {Gait and speed
  selection in slender inertial swimmers},\ }\href@noop {} {\bibfield
  {journal} {\bibinfo  {journal} {Proceedings of the National Academy of
  Sciences}\ }\textbf {\bibinfo {volume} {112}},\ \bibinfo {pages} {3874}
  (\bibinfo {year} {2015})}\BibitemShut {NoStop}%
\bibitem [{\citenamefont {Van~Buren}\ \emph {et~al.}(2018)\citenamefont
  {Van~Buren}, \citenamefont {Floryan}, \citenamefont {Wei},\ and\
  \citenamefont {Smits}}]{VanBuren2018}%
  \BibitemOpen
  \bibfield  {author} {\bibinfo {author} {\bibfnamefont {T.}~\bibnamefont
  {Van~Buren}}, \bibinfo {author} {\bibfnamefont {D.}~\bibnamefont {Floryan}},
  \bibinfo {author} {\bibfnamefont {N.}~\bibnamefont {Wei}},\ and\ \bibinfo
  {author} {\bibfnamefont {A.~J.}\ \bibnamefont {Smits}},\ }\bibfield  {title}
  {\bibinfo {title} {Flow speed has little impact on propulsive characteristics
  of oscillating foils},\ }\href@noop {} {\bibfield  {journal} {\bibinfo
  {journal} {Physical Review Fluids}\ }\textbf {\bibinfo {volume} {3}},\
  \bibinfo {pages} {013103} (\bibinfo {year} {2018})}\BibitemShut {NoStop}%
\bibitem [{\citenamefont {Guyon}\ \emph {et~al.}(2015)\citenamefont {Guyon},
  \citenamefont {Hulin}, \citenamefont {Petit},\ and\ \citenamefont
  {Mitescu}}]{GuyonHulinPetit}%
  \BibitemOpen
  \bibfield  {author} {\bibinfo {author} {\bibfnamefont {E.}~\bibnamefont
  {Guyon}}, \bibinfo {author} {\bibfnamefont {J.-P.}\ \bibnamefont {Hulin}},
  \bibinfo {author} {\bibfnamefont {L.}~\bibnamefont {Petit}},\ and\ \bibinfo
  {author} {\bibfnamefont {C.~D.}\ \bibnamefont {Mitescu}},\ }\href@noop {}
  {\emph {\bibinfo {title} {Physical Hydrodynamics, 2nd Edition}}}\ (\bibinfo
  {publisher} {Oxford, University Press},\ \bibinfo {year} {2015})\BibitemShut
  {NoStop}%
\bibitem [{\citenamefont {Brennen}(2006)}]{BrennenBook}%
  \BibitemOpen
  \bibfield  {author} {\bibinfo {author} {\bibfnamefont {C.~E.}\ \bibnamefont
  {Brennen}},\ }\href {http://brennen.caltech.edu/fluidbook/FLUIDBOOK.htm}
  {\emph {\bibinfo {title} {An Internet Book on Fluid Dynamics}}}\ (\bibinfo
  {publisher} {Dankat Publishing.},\ \bibinfo {year} {2006})\BibitemShut
  {NoStop}%
\bibitem [{\citenamefont {Brennen}(1982)}]{BrennenReview}%
  \BibitemOpen
  \bibfield  {author} {\bibinfo {author} {\bibfnamefont {C.~E.}\ \bibnamefont
  {Brennen}},\ }\href@noop {} {\emph {\bibinfo {title} {A review of added mass
  and fluid inertial forces}}},\ \bibinfo {type} {Tech. Rep.}\ (\bibinfo
  {institution} {aval Civil Eng. Lab., Port Hueneme, Calif., Report CR82.010},\
  \bibinfo {year} {1982})\BibitemShut {NoStop}%
\bibitem [{\citenamefont {Sharma}\ \emph {et~al.}(2022)\citenamefont {Sharma},
  \citenamefont {Berhanu},\ and\ \citenamefont {Kudrolli}}]{Sharma2021}%
  \BibitemOpen
  \bibfield  {author} {\bibinfo {author} {\bibfnamefont {R.~S.}\ \bibnamefont
  {Sharma}}, \bibinfo {author} {\bibfnamefont {M.}~\bibnamefont {Berhanu}},\
  and\ \bibinfo {author} {\bibfnamefont {A.}~\bibnamefont {Kudrolli}},\
  }\bibfield  {title} {\bibinfo {title} {Alcove formation in dissolving cliffs
  driven by density inversion instability},\ }\href
  {https://doi.org/10.1063/5.0092331} {\bibfield  {journal} {\bibinfo
  {journal} {Physics of Fluids}\ }\textbf {\bibinfo {volume} {34}},\ \bibinfo
  {pages} {054118} (\bibinfo {year} {2022})}\BibitemShut {NoStop}%
\bibitem [{\citenamefont {Pegler}\ and\ \citenamefont
  {{Davies~Wykes}}(2020)}]{Pegler2020}%
  \BibitemOpen
  \bibfield  {author} {\bibinfo {author} {\bibfnamefont {S.~S.}\ \bibnamefont
  {Pegler}}\ and\ \bibinfo {author} {\bibfnamefont {M.~S.}\ \bibnamefont
  {{Davies~Wykes}}},\ }\bibfield  {title} {\bibinfo {title} {{Shaping of
  melting and dissolving solids under natural convection}},\ }\href
  {https://doi.org/10.1017/jfm.2020.507} {\bibfield  {journal} {\bibinfo
  {journal} {J. Fluid Mech.}\ }\textbf {\bibinfo {volume} {900}},\ \bibinfo
  {pages} {A35} (\bibinfo {year} {2020})}\BibitemShut {NoStop}%
\bibitem [{\citenamefont {Chandrasekhar}(1961)}]{Chandrasekhar}%
  \BibitemOpen
  \bibfield  {author} {\bibinfo {author} {\bibfnamefont {S.}~\bibnamefont
  {Chandrasekhar}},\ }\href@noop {} {\emph {\bibinfo {title} {Hydrodynamic and
  Hydromagnetic Stability}}}\ (\bibinfo  {publisher} {Clarendon Press,
  Oxford},\ \bibinfo {year} {1961})\BibitemShut {NoStop}%
\bibitem [{\citenamefont {Schlichting}(1979)}]{Schlichting}%
  \BibitemOpen
  \bibfield  {author} {\bibinfo {author} {\bibfnamefont {H.}~\bibnamefont
  {Schlichting}},\ }\href@noop {} {\emph {\bibinfo {title} {Boundary layer
  theory}}}\ (\bibinfo  {publisher} {McGraw-Hill},\ \bibinfo {year}
  {1979})\BibitemShut {NoStop}%
\bibitem [{\citenamefont {Meakin}\ and\ \citenamefont
  {Jamtveit}(2010)}]{Meakin2010}%
  \BibitemOpen
  \bibfield  {author} {\bibinfo {author} {\bibfnamefont {P.}~\bibnamefont
  {Meakin}}\ and\ \bibinfo {author} {\bibfnamefont {B.}~\bibnamefont
  {Jamtveit}},\ }\bibfield  {title} {\bibinfo {title} {Geological pattern
  formation by growth and dissolution in aqueous systems},\ }\href
  {https://doi.org/10.1098/rspa.2009.0189} {\bibfield  {journal} {\bibinfo
  {journal} {Proc. R. Soc. A}\ }\textbf {\bibinfo {volume} {466}},\ \bibinfo
  {pages} {659–694} (\bibinfo {year} {2010})}\BibitemShut {NoStop}%
\bibitem [{\citenamefont {Cohen}\ \emph {et~al.}(2016)\citenamefont {Cohen},
  \citenamefont {Berhanu}, \citenamefont {Derr},\ and\ \citenamefont
  {Courrech~du Pont}}]{Cohen2016}%
  \BibitemOpen
  \bibfield  {author} {\bibinfo {author} {\bibfnamefont {C.}~\bibnamefont
  {Cohen}}, \bibinfo {author} {\bibfnamefont {M.}~\bibnamefont {Berhanu}},
  \bibinfo {author} {\bibfnamefont {J.}~\bibnamefont {Derr}},\ and\ \bibinfo
  {author} {\bibfnamefont {S.}~\bibnamefont {Courrech~du Pont}},\ }\bibfield
  {title} {\bibinfo {title} {Erosion patterns on dissolving and melting bodies
  (2015 gallery of fluid motion)},\ }\href
  {https://doi.org/10.1103/PhysRevFluids.1.050508} {\bibfield  {journal}
  {\bibinfo  {journal} {Phys. Rev. Fluids}\ }\textbf {\bibinfo {volume} {1}},\
  \bibinfo {pages} {050508} (\bibinfo {year} {2016})}\BibitemShut {NoStop}%
\bibitem [{\citenamefont {Kerr}(1994{\natexlab{a}})}]{KerrJFM1994b}%
  \BibitemOpen
  \bibfield  {author} {\bibinfo {author} {\bibfnamefont {R.~C.}\ \bibnamefont
  {Kerr}},\ }\bibfield  {title} {\bibinfo {title} {Melting driven by vigorous
  compositional convection},\ }\href
  {https://doi.org/10.1017/S0022112094002922} {\bibfield  {journal} {\bibinfo
  {journal} {Journal of Fluid Mechanics}\ }\textbf {\bibinfo {volume} {280}},\
  \bibinfo {pages} {255} (\bibinfo {year} {1994}{\natexlab{a}})}\BibitemShut
  {NoStop}%
\bibitem [{\citenamefont {Kerr}(1994{\natexlab{b}})}]{KerrJFM1994}%
  \BibitemOpen
  \bibfield  {author} {\bibinfo {author} {\bibfnamefont {R.~C.}\ \bibnamefont
  {Kerr}},\ }\bibfield  {title} {\bibinfo {title} {Dissolving driven by
  vigorous compositional convection},\ }\href
  {https://doi.org/10.1017/S0022112094002934} {\bibfield  {journal} {\bibinfo
  {journal} {Journal of Fluid Mechanics}\ }\textbf {\bibinfo {volume} {280}},\
  \bibinfo {pages} {287} (\bibinfo {year} {1994}{\natexlab{b}})}\BibitemShut
  {NoStop}%
\bibitem [{\citenamefont {Hewitt}(2020)}]{Hewitt2020}%
  \BibitemOpen
  \bibfield  {author} {\bibinfo {author} {\bibfnamefont {I.~J.}\ \bibnamefont
  {Hewitt}},\ }\bibfield  {title} {\bibinfo {title} {Subglacial plumes},\
  }\href {https://doi.org/10.1146/annurev-fluid-010719-060252} {\bibfield
  {journal} {\bibinfo  {journal} {Annual Review of Fluid Mechanics}\ }\textbf
  {\bibinfo {volume} {52}},\ \bibinfo {pages} {145} (\bibinfo {year}
  {2020})}\BibitemShut {NoStop}%
\bibitem [{\citenamefont {McConnochie}\ and\ \citenamefont
  {Kerr}(2016)}]{mcconnochie_kerr_2016}%
  \BibitemOpen
  \bibfield  {author} {\bibinfo {author} {\bibfnamefont {C.~D.}\ \bibnamefont
  {McConnochie}}\ and\ \bibinfo {author} {\bibfnamefont {R.~C.}\ \bibnamefont
  {Kerr}},\ }\bibfield  {title} {\bibinfo {title} {The turbulent wall plume
  from a vertically distributed source of buoyancy},\ }\href
  {https://doi.org/10.1017/jfm.2015.691} {\bibfield  {journal} {\bibinfo
  {journal} {Journal of Fluid Mechanics}\ }\textbf {\bibinfo {volume} {787}},\
  \bibinfo {pages} {237–253} (\bibinfo {year} {2016})}\BibitemShut {NoStop}%
\bibitem [{\citenamefont {McKenna}(2005)}]{mckenna2005}%
  \BibitemOpen
  \bibfield  {author} {\bibinfo {author} {\bibfnamefont {R.}~\bibnamefont
  {McKenna}},\ }\bibfield  {title} {\bibinfo {title} {Iceberg shape
  characterization},\ }in\ \href@noop {} {\emph {\bibinfo {booktitle}
  {Proceedings 18th International Conference on Port and Ocean Engineering
  under Arctic Conditions}}},\ Vol.~\bibinfo {volume} {2}\ (\bibinfo {year}
  {2005})\ pp.\ \bibinfo {pages} {555--564}\BibitemShut {NoStop}%
\bibitem [{\citenamefont {Cenedese}\ and\ \citenamefont
  {Straneo}(2023)}]{cenedese2023}%
  \BibitemOpen
  \bibfield  {author} {\bibinfo {author} {\bibfnamefont {C.}~\bibnamefont
  {Cenedese}}\ and\ \bibinfo {author} {\bibfnamefont {F.}~\bibnamefont
  {Straneo}},\ }\bibfield  {title} {\bibinfo {title} {Icebergs melting},\
  }\href@noop {} {\bibfield  {journal} {\bibinfo  {journal} {Annual Review of
  Fluid Mechanics}\ }\textbf {\bibinfo {volume} {55}} (\bibinfo {year}
  {2023})}\BibitemShut {NoStop}%
\bibitem [{\citenamefont {Romanov}\ \emph {et~al.}(2012)\citenamefont
  {Romanov}, \citenamefont {Romanova},\ and\ \citenamefont
  {Romanov}}]{romanov2012}%
  \BibitemOpen
  \bibfield  {author} {\bibinfo {author} {\bibfnamefont {Y.~A.}\ \bibnamefont
  {Romanov}}, \bibinfo {author} {\bibfnamefont {N.~A.}\ \bibnamefont
  {Romanova}},\ and\ \bibinfo {author} {\bibfnamefont {P.}~\bibnamefont
  {Romanov}},\ }\bibfield  {title} {\bibinfo {title} {Shape and size of
  antarctic icebergs derived from ship observation data},\ }\href@noop {}
  {\bibfield  {journal} {\bibinfo  {journal} {Antarctic Science}\ }\textbf
  {\bibinfo {volume} {24}},\ \bibinfo {pages} {77} (\bibinfo {year}
  {2012})}\BibitemShut {NoStop}%
\bibitem [{\citenamefont {Weady}\ \emph {et~al.}(2022)\citenamefont {Weady},
  \citenamefont {Tong}, \citenamefont {Zidovska},\ and\ \citenamefont
  {Ristroph}}]{Weady2022}%
  \BibitemOpen
  \bibfield  {author} {\bibinfo {author} {\bibfnamefont {S.}~\bibnamefont
  {Weady}}, \bibinfo {author} {\bibfnamefont {J.}~\bibnamefont {Tong}},
  \bibinfo {author} {\bibfnamefont {A.}~\bibnamefont {Zidovska}},\ and\
  \bibinfo {author} {\bibfnamefont {L.}~\bibnamefont {Ristroph}},\ }\bibfield
  {title} {\bibinfo {title} {Anomalous convective flows carve pinnacles and
  scallops in melting ice},\ }\href@noop {} {\bibfield  {journal} {\bibinfo
  {journal} {Physical Review Letters}\ }\textbf {\bibinfo {volume} {128}},\
  \bibinfo {pages} {044502} (\bibinfo {year} {2022})}\BibitemShut {NoStop}%
\bibitem [{\citenamefont {Bohleber}\ \emph {et~al.}(2017)\citenamefont
  {Bohleber}, \citenamefont {Sold}, \citenamefont {Hardy}, \citenamefont
  {Schwikowski}, \citenamefont {Klenk}, \citenamefont {Fischer}, \citenamefont
  {Sirguey}, \citenamefont {Cullen}, \citenamefont {Potocki}, \citenamefont
  {Hoffmann},\ and\ \citenamefont {Mayewski}}]{Bohleber2017}%
  \BibitemOpen
  \bibfield  {author} {\bibinfo {author} {\bibfnamefont {P.}~\bibnamefont
  {Bohleber}}, \bibinfo {author} {\bibfnamefont {L.}~\bibnamefont {Sold}},
  \bibinfo {author} {\bibfnamefont {D.~R.}\ \bibnamefont {Hardy}}, \bibinfo
  {author} {\bibfnamefont {M.}~\bibnamefont {Schwikowski}}, \bibinfo {author}
  {\bibfnamefont {P.}~\bibnamefont {Klenk}}, \bibinfo {author} {\bibfnamefont
  {A.}~\bibnamefont {Fischer}}, \bibinfo {author} {\bibfnamefont
  {P.}~\bibnamefont {Sirguey}}, \bibinfo {author} {\bibfnamefont {N.~J.}\
  \bibnamefont {Cullen}}, \bibinfo {author} {\bibfnamefont {M.}~\bibnamefont
  {Potocki}}, \bibinfo {author} {\bibfnamefont {H.}~\bibnamefont {Hoffmann}},\
  and\ \bibinfo {author} {\bibfnamefont {P.}~\bibnamefont {Mayewski}},\
  }\bibfield  {title} {\bibinfo {title} {Ground-penetrating radar reveals ice
  thickness and undisturbed englacial layers at kilimanjaro's northern ice
  field},\ }\href {https://doi.org/10.5194/tc-11-469-2017} {\bibfield
  {journal} {\bibinfo  {journal} {The Cryosphere}\ }\textbf {\bibinfo {volume}
  {11}},\ \bibinfo {pages} {469} (\bibinfo {year} {2017})}\BibitemShut
  {NoStop}%
\bibitem [{\citenamefont {Zhou}\ \emph {et~al.}(2019)\citenamefont {Zhou},
  \citenamefont {Bachmayer},\ and\ \citenamefont {deYoung}}]{Zhou2019}%
  \BibitemOpen
  \bibfield  {author} {\bibinfo {author} {\bibfnamefont {M.}~\bibnamefont
  {Zhou}}, \bibinfo {author} {\bibfnamefont {R.}~\bibnamefont {Bachmayer}},\
  and\ \bibinfo {author} {\bibfnamefont {B.}~\bibnamefont {deYoung}},\
  }\bibfield  {title} {\bibinfo {title} {Mapping the underside of an iceberg
  with a modified underwater glider},\ }\href
  {https://doi.org/https://doi.org/10.1002/rob.21873} {\bibfield  {journal}
  {\bibinfo  {journal} {Journal of Field Robotics}\ }\textbf {\bibinfo {volume}
  {36}},\ \bibinfo {pages} {1102} (\bibinfo {year} {2019})},\ \Eprint
  {https://arxiv.org/abs/https://onlinelibrary.wiley.com/doi/pdf/10.1002/rob.21873}
  {https://onlinelibrary.wiley.com/doi/pdf/10.1002/rob.21873} \BibitemShut
  {NoStop}%
\bibitem [{\citenamefont {Wagner}\ \emph {et~al.}(2017)\citenamefont {Wagner},
  \citenamefont {Dell},\ and\ \citenamefont {Eisenman}}]{Wagner2017}%
  \BibitemOpen
  \bibfield  {author} {\bibinfo {author} {\bibfnamefont {T.~J.}\ \bibnamefont
  {Wagner}}, \bibinfo {author} {\bibfnamefont {R.~W.}\ \bibnamefont {Dell}},\
  and\ \bibinfo {author} {\bibfnamefont {I.}~\bibnamefont {Eisenman}},\
  }\bibfield  {title} {\bibinfo {title} {An analytical model of iceberg
  drift},\ }\href {https://doi.org/10.1175/JPO-D-16-0262.1} {\bibfield
  {journal} {\bibinfo  {journal} {Journal of Physical Oceanography}\ }\textbf
  {\bibinfo {volume} {47}},\ \bibinfo {pages} {1605} (\bibinfo {year}
  {2017})}\BibitemShut {NoStop}%
\bibitem [{\citenamefont {Marchenko}\ \emph {et~al.}(2019)\citenamefont
  {Marchenko}, \citenamefont {Diansky},\ and\ \citenamefont
  {Fomin}}]{Marchenko2019}%
  \BibitemOpen
  \bibfield  {author} {\bibinfo {author} {\bibfnamefont {A.}~\bibnamefont
  {Marchenko}}, \bibinfo {author} {\bibfnamefont {N.}~\bibnamefont {Diansky}},\
  and\ \bibinfo {author} {\bibfnamefont {V.}~\bibnamefont {Fomin}},\ }\bibfield
   {title} {\bibinfo {title} {Modeling of iceberg drift in the marginal ice
  zone of the barents sea},\ }\href
  {https://doi.org/https://doi.org/10.1016/j.apor.2019.03.008} {\bibfield
  {journal} {\bibinfo  {journal} {Applied Ocean Research}\ }\textbf {\bibinfo
  {volume} {88}},\ \bibinfo {pages} {210} (\bibinfo {year} {2019})}\BibitemShut
  {NoStop}%
\bibitem [{\citenamefont {Liberzon}\ \emph {et~al.}(2021)\citenamefont
  {Liberzon}, \citenamefont {Käufer}, \citenamefont {Bauer}, \citenamefont
  {Vennemann},\ and\ \citenamefont {Zimmer}}]{OpenPIV}%
  \BibitemOpen
  \bibfield  {author} {\bibinfo {author} {\bibfnamefont {A.}~\bibnamefont
  {Liberzon}}, \bibinfo {author} {\bibfnamefont {T.}~\bibnamefont {Käufer}},
  \bibinfo {author} {\bibfnamefont {A.}~\bibnamefont {Bauer}}, \bibinfo
  {author} {\bibfnamefont {P.}~\bibnamefont {Vennemann}},\ and\ \bibinfo
  {author} {\bibfnamefont {E.}~\bibnamefont {Zimmer}},\ }\href
  {https://doi.org/10.5281/zenodo.4409178} {\bibinfo {title}
  {Openpiv/openpiv-python: Openpiv-python v0.23.4}} (\bibinfo {year}
  {2021})\BibitemShut {NoStop}%
\end{thebibliography}
\end{document}